\documentclass{emulateapj}

\newcommand{\Msun}{M_\odot}
\newcommand{\sci}[1]{\times10^{#1}}
\newcommand{\fig}[1]{Figure \ref{fig:#1}}
\newcommand{\figs}[2]{Figures \ref{fig:#1} and \ref{fig:#2}}
\newcommand{\figsto}[2]{Figures \ref{fig:#1}--\ref{fig:#2}}
\newcommand{\tab}[1]{Table \ref{table:#1}}
\newcommand{\eqn}[1]{Equation (\ref{eqn:#1})}

\newcommand{\lcdm}{$\Lambda$CDM}
\newcommand{\kms}{km\,s$^{-1}$}
\newcommand{\mathkms}{\mathrm{km}\,\mathrm{s}^{-1}}
\newcommand{\degrees}{^\circ}
\newcommand{\newrevision}{}
\newcommand{\newrevisionII}{}
\shorttitle{Using stream gaps to probe missing satellites}
\shortauthors{Ngan \& Carlberg}

\begin{document}

%% LaTeX will automatically break titles if they run longer than
%% one line. However, you fmay use \\ to force a line break if
%% you desire.

\title{Using gaps in N-body tidal streams to probe missing satellites}
%\title{Probing missing satellites using N-body simulations of tidal streams}
%\title{Probing the abundance of missing satellites using density gaps in tidal streams}
% \title{Simulating density gaps in tidal streams using missing satellites}

%% Use \author, \affil, and the \and command to format
%% author and affiliation information.
%% Note that \email has replaced the old \authoremail command
%% from AASTeX v4.0. You can use \email to mark an email address
%% anywhere in the paper, not just in the front matter.
%% As in the title, use \\ to force line breaks.

\author{W. H. W. Ngan and R. G. Carlberg}
\affil{Department of Astronomy and Astrophysics, University of Toronto, Toronto, ON, M5S3H4, Canada}
\email{ngan@astro.utoronto.ca}

% \author{S. Djorgovski\altaffilmark{1,2,3} and Ivan R. King\altaffilmark{1}}
% \affil{Astronomy Department, University of California,
%     Berkeley, CA 94720}
% 
% \author{C. D. Biemesderfer\altaffilmark{4,5}}
% \affil{National Optical Astronomy Observatories, Tucson, AZ 85719}
% \email{aastex-help@aas.org}
% 
% \and
% 
% \author{R. J. Hanisch\altaffilmark{5}}
% \affil{Space Telescope Science Institute, Baltimore, MD 21218}

%% Notice that each of these authors has alternate affiliations, which
%% are identified by the \altaffilmark after each name.  Specify alternate
%% affiliation information with \altaffiltext, with one command per each
%% affiliation.

% \altaffiltext{1}{Visiting Astronomer, Cerro Tololo Inter-American Observatory.
% CTIO is operated by AURA, Inc.\ under contract to the National Science
% Foundation.}
% \altaffiltext{2}{Society of Fellows, Harvard University.}
% \altaffiltext{3}{present address: Center for Astrophysics,
%     60 Garden Street, Cambridge, MA 02138}
% \altaffiltext{4}{Visiting Programmer, Space Telescope Science Institute}
% \altaffiltext{5}{Patron, Alonso's Bar and Grill}

%% Mark off your abstract in the ``abstract'' environment. In the manuscript
%% style, abstract will output a Received/Accepted line after the
%% title and affiliation information. No date will appear since the author
%% does not have this information. The dates will be filled in by the
%% editorial office after submission.

\begin{abstract}

We use N-body simulations to model the tidal disruption of a star cluster in a Milky-Way-sized dark matter halo,
which results in a narrow stream comparable to (but slightly wider than) Pal-5 or GD-1.
The mean Galactic dark matter halo is modeled by a spherical
Navarro-Frenk-White (NFW) potential with subhalos predicted by the \lcdm\ cosmological model. The distribution and mass function of
the subhalos follow the results from the Aquarius simulation. We use a matched filter approach
to look for ``gaps'' in tidal streams at 12 length scales from $0.1\,\mathrm{kpc}$ to $5\,\mathrm{kpc}$,
which appear as characteristic dips in the linear densities along the streams.
We find that, in addition to the subhalos' perturbations, the epicyclic overdensities (EOs)
due to the coherent epicyclic motions of particles in a stream also produce gap-like 
signals near the progenitor. We measure the gap spectra -- the gap formation rates as functions of gap length --
due to both subhalo perturbations and EOs, which have not been accounted for together by
previous studies. \newrevision{Finally, we project the simulated streams onto the sky to investigate issues
when interpreting gap spectra in observations. In particular, we find that gap spectra from low signal-to-noise
observations can be biased by the orbital phase of the stream. This indicates that the study of stream gaps
will benefit greatly from high-quality data from future missions.
}

%When compared to the observed gap spectrum of GD-1, we conclude that GD-1's gaps
%are inconsistent with EOs alone in a smooth dark matter halo, but are roughly consistent with the 
%presence of \lcdm\ subhalos.

%We use N-body simulations to simulate the formation of a tidal stream similar to GD-1 in a Milky Way sized
%dark matter halo, which is modelled by a spherical NFW potential and subhalos
%ranging from $10^6\Msun$ to $10^8\Msun$. The distribution and mass function of
%subhalos follow the results from the Aquarius simulation \citep{aquarius}. We divide
%the set of subhalos into six mass ranges, and we compute the linear densities along
%the stream. We find that the subhalos leave significant perturbations on the stream
%which can be seen as dips in its density profile. This suggest that dark matter subhalos
%can be the origin of the density gaps that were previously observed in tidal streams.
\end{abstract}

%% Keywords should appear after the \end{abstract} command. The uncommented
%% example has been keyed in ApJ style. See the instructions to authors
%% for the journal to which you are submitting your paper to determine
%% what keyword punctuation is appropriate.

%\keywords{Tidal streams, subhalos, LOOK UP HOW TO WRITE THIS LIST}

% \keywords{globular clusters: general --- globular clusters: individual(NGC 6397,
% NGC 6624, NGC 7078, Terzan 8}

%% From the front matter, we move on to the body of the paper.
%% In the first two sections, notice the use of the natbib \citep
%% and \citet commands to identify citations.  The citations are
%% tied to the reference list via symbolic KEYs. The KEY corresponds
%% to the KEY in the \bibitem in the reference list below. We have
%% chosen the first three characters of the first author's name plus
%% the last two numeral of the year of publication as our KEY for
%% each reference.

%% Authors who wish to have the most important objects in their paper
%% linked in the electronic edition to a data center may do so by tagging
%% their objects with \objectname{} or \object{}.  Each macro takes the
%% object name as its required argument. The optional, square-bracket 
%% argument should be used in cases where the data center identification
%% differs from what is to be printed in the paper.  The text appearing 
%% in curly braces is what will appear in print in the published paper. 
%% If the object name is recognized by the data centers, it will be linked
%% in the electronic edition to the object data available at the data centers  
%%
%% Note that for sources with brackets in their names, e.g. [WEG2004] 14h-090,
%% the brackets must be escaped with backslashes when used in the first
%% square-bracket argument, for instance, \object[\[WEG2004\] 14h-090]{90}).
%%  Otherwise, LaTeX will issue an error. 

\section{Introduction}

The \lcdm\ model -- a universe dominated by cold dark matter and a cosmological constant -- is
a successful model of the universe at large scales
\citep{blumenthal84, davis85, bardeen86, riess98, perlmutter99, planck13}. However, as more observations are made and computational
power becomes available, discrepancies in the model have been found below galactic scales.
Cosmological simulations show that the dark matter halo of a Milky-Way-sized galaxy,
in addition to having a smooth component with well-known density profiles \citep{nfw, navarro04, navarro10}, should be populated
by substructures, or subhalos \citep{madau08, stadel09, zemp09, aquarius, gao11}.
This theoretical prediction has not been supported by
observational evidence, as many efforts over the past two decades have failed to find
enough satellite systems in the Milky Way to account for the predicted abundance \citep{klypin99, moore99, strigari07}. This discrepancy is called the
{\it Missing Satellite Problem}.

Competing solutions to the Missing Satellite Problem can be roughly classified into two types --
astrophysical and physical. Astrophysical solutions, on one hand, postulate that the predicted subhalo
abundance is correct, but the subhalos have too little stellar
content to be observable directly. For example, reionizing radiation or stellar feedback \citep{koposov09, maccio09}
can suppress star formation in a subhalo. This class of solutions \newrevision{is} founded on the absence of
baryonic processes in those cosmological simulations which predicted the subhalo abundances.
Physical solutions, on the other hand, postulate that the predicted subhalo abundance is
incorrect, as our understanding of CDM may be incomplete. Alternative dark matter
solutions, such as warm dark matter \citep{barkana01, bode01, benson13, schneider13}, self-interacting dark
matter \citep{spergel00}, or inflationary models with non-scale invariance \citep{kamionkowski00} offer mechanisms
to suppress structure formation at small scales. Clearly, the predicted difference
between these two types of solutions is the number of subhalos. In order to test the \lcdm\ model at
sub-galactic scales, measuring the true abundance of subhalos is an important step.

Tidal streams -- or simply ``streams'' -- are remnants of stellar systems such as globular clusters (GCs) or dwarf galaxies (DGs) as 
they are tidally disrupted by a massive host. When the stars become unbound from the progenitor, the stars trace an
elongated tail which wraps around the massive host. Tidal remnants have long been useful probes for studying the gravitational potential 
of the Milky Way \citep[e.g.,][]{johnston98, helmi03, law05}. In particular, \citet{ibata02} first used simulations to show that
the encounters between the stream stars and subhalos can dynamically heat up the stream, which can be used to probe the
presence of subhalos. Moreover, a key influence that subhalos have on streams is that the stream stars near the point
of the encounter get scattered into different orbits by the perturbation,
causing an abrupt decrease in stellar density in that region of an otherwise smooth stream. Using the abundance of subhalos obtained
from high-resolution simulations \citep{madau08, aquarius},
it has been predicted that streams in the Milky Way described by the \lcdm\ model should contain many ``gaps''
\citep{yoon11, carlberg12, pal5_gaps, CG13}.

In the past decade, many streams in the Milky Way have been found (see \citet{G10} for a list).
Two streams of particular interest to us are Pal-5 \citep{pal5} and GD-1 \citep{gd1}.
Both streams are detected as long, narrow tidal tails with length-to-width ratio of $\sim100$.
These two streams are interesting as they show varying densities longitudinally along the streams.
It is not clear whether those density variations correspond to subhalo perturbations.
Other possible origins of those density variations include clumping due to the coherence in the
epicyclic orbits of stream stars \citep{kupper08, kupper10} and Jeans instabilities \citep{qc11}. The ultimate goal of
this study is to test whether the gaps observed in streams
are consistent with the prediction by the \lcdm\ model. \citet{yoon11} and \citet{carlberg12} independently
made the first predictions by simulating ideal streams with massless particles in the presence of orbiting subhalos.
In particular, \citet{carlberg12} derived analytical expressions of the gap formation rate as
a function of a stream's intrinsic properties, which are readily comparable to observations. However, neither of the aforementioned
studies self-consistently modeled the realistic disruption of the progenitor system. 

In this study, we measure the gap formation rate by modeling
a stream's formation and its interaction with subhalos using N-body simulations. This paper is organized as follows.
Section 2 describes the details of our simulations, including the subhalo
abundances and density profiles that we adapt, and the details of the star cluster and the resulting model stream.
Section 3 focuses on the method of detecting gaps in a simulated stream. The
method of using match filters is inspired by analyses for observations, but modified here to analyze
simulations. Section 4 contains detailed discussion of our key results, including the phenomenology of gaps and comparisons
with previous analytical predictions and observations. Section 5 is a summary of our results.

% \begin{itemize}

% \item Missing satellites and $\Lambda$CDM cosmology. There are some solutions like baryonic effects
% and warm dark matter that can solve problem, but still need observational evidence for 
% missing satellites.

% \item Lots of tidal streams have been observed. Widely used as probe for galactic potential, but
% density fluctuations in some streams like Pal-5 and GD-1 are not well explained. Maybe
% epicyclic clumps \citep{kupper08}, instabilities \citep{qc11}. but recently been used to study subhalos
% \citep{yoon11, carlberg12}.

% \item \citet{carlberg12} derived gap formation rates semi-analycially using idealized particles.
% \citet{yoon11} looked at general morphology of tidal streams bombarded by different subhalo masses.
% But neither used N-body to get realistic density profiles which results from disruption of a satellite
% system.

% \item Introduce GD-1 \citep{gd1} and Pal-5 \citep{pal5}. Pal-5 is closer to Galactic center so suffers more disk-shocking. GD-1
% has peri- and apogalactic distances at about 15 kpc and 30 kpc \citep{willett09}, so don't have to worry
% about disk as much. More facts about GD-1 like width and velocity dispersion. We will use GD-1 as a
% model stream.

% \end{itemize}

\section{Simulations}

\subsection{Models}
\label{sec:models}

The host galaxy is modeled as a dark matter halo, as well as a set of subhalos which orbit
around the halo's potential. A Milky-Way-sized halo is modeled with a static spherical Navarro-Frenk-White (NFW) profile
\citep{nfw} with $v_{max}=210$ \kms\ located at $r_{max}=30\,\mathrm{kpc}$.
% \begin{equation}
%     \Phi_{gal}(r) = -4\pi G \rho_0 a^2 \frac{\ln(1+r/a)}{r/a}.
% \end{equation}
% where $\rho_0 = ?$ and $a=?$. 
Each individual subhalo is modeled by a spherical Hernquist profile \citep{hernquist90} 
\begin{equation}
    \Phi_i(r) = \frac{GM_i}{h_i+r}
\end{equation}
for simplicity, compared to Einasto profiles which produce better fits in simulations but are more complicated to compute
\citep{aquarius}. We use the \newrevision{formula found in \citet{carlberg09}, which approximates the results from both
\citet{aquarius} and \citet{neto07},} where $h_i$ is independent of galactocentric position,
and is related to $M_i$ by
\begin{equation}
    h(M) = 6\,\mathrm{kpc} \times \left( \frac{M}{10^{10}\Msun} \right)^{0.43}.
    \label{eqn:subhalo_size}
\end{equation}
We use the mass and spatial distributions of the subhalos from the results of the Aquarius simulations \citep{aquarius},
where the mass function is independent of the spatial distribution.
The mass function is a power law
\begin{equation}
    \frac{dN}{dM} = 3.26\sci{-5}\Msun^{-1} \left( \frac{M}{2.52\sci{7}\Msun} \right)^{-1.9},
    \label{eqn:subhalo_massfunction}
\end{equation}
% where $n=-1.9$, $a_0=3.26\sci{-5}\Msun^{-1}$, and $m_0=2.52\sci{7}\Msun$.
and the spatial distribution follows an Einasto profile
\begin{equation}
    n(r) \propto \exp \left\{ -2.95 \left[ \left(\frac{r}{199\,\mathrm{kpc}} \right)^{0.678} - 1 \right]  \right\}.
\end{equation}
The subhalos' velocities are initialized with a Gaussian distribution where the velocity dispersion is the solution to
the isotropic Jean's equation \citep{bt08} using the halo's 
potential. The subhalos orbit around this potential as test masses.

The progenitor of the stream, which is an approximation to a globular star cluster,
is initialized using $10^6$ particles of equal mass as a King model with parameters $w=4.91$, total mass $4.29\sci{4}\Msun$, and a
core radius of $0.01$ kpc. This results in a zero-density radius of $0.103$ kpc.
Each N-body particle in the system interacts with the \newrevision{dark matter halo's and subhalos' potentials.}
With the Galactic center at the origin, the satellite is initially put at $(x,y,z)=(30, 0, 0)$ kpc and
velocity $(v_x, v_y, v_z) = (0, 140, 0)$ \kms. 
The resulting orbit is confined on the $xy$-plane with eccentricity $0.33$, peri- and apogalacticon at
$r_p=15\,\mathrm{kpc}$ and $r_a=30\,\mathrm{kpc}$,
respectively. The azimuthal and radial periods are about 0.70 Gyr and 0.47 Gyr, respectively.

%, shown in \fig{stream_orbit}.

%\begin{figure}
%    \centering
%    \includegraphics[width=6in]{figs/stream_8Gyr/stream_8Gyr.eps}
%    \caption{The smooth stream at 8 Gyr (black line) and the satellite's orbit from 6 to 8 Gyr
%    (gray dotted line). The stream orbits in the $xy$-plane in counter-clockwise direction as viewed from the $+z$ position.
%    A detailed view of the stream is shown in the top panel of \fig{diagonal_gap_map}.}
%    \label{fig:stream_orbit}
%\end{figure}

% is in rough agreement with
%previous observations of GD-1 . 

%in an
%eccentric orbit ($e=0.33$ in our simulation), so both the width and velocity dispersion vary along the stream. 
%The observed part of GD-1 has velocity dispersion of $\lesssim 3$ km/s and width of $< 100$ pc, so our simulated
%stream is a few times wider than the observed stream in some regions. Gaps in a wider stream can suffer from
%shearing due to different angular frequencies along the width of the stream (**maybe a reference here**), so 
%gaps are more easily erased in a wide stream than in a narrow stream. Therefore, having a wider stream in our
%simulation should not affect our conclusion.

\subsection{Software and Parameters}

We use {\sc Gadget-2} \citep{gadget2} for our N-body simulations. Since the public
distribution\footnote{\url{http://www.mpa-garching.mpg.de/gadget/}}
does not have functionality for external potentials, we modify the code such that in every time step,
an external acceleration term which accounts for the potentials of the halo and all the subhalos is added
to the accelerations of all the particles after their N-body 
interactions are computed.

Each of our simulations lasts 10 Gyr, and we impose
a maximum time step of 1 Myr. The particle softening is 5 pc. Each simulation produces 500 snapshots, one every
20 Myr, and each consists of the positions and velocities of the N-body particles and subhalos.

\subsection{Stream Properties}

\begin{figure}
	\centering
	\includegraphics[width=3.5in]{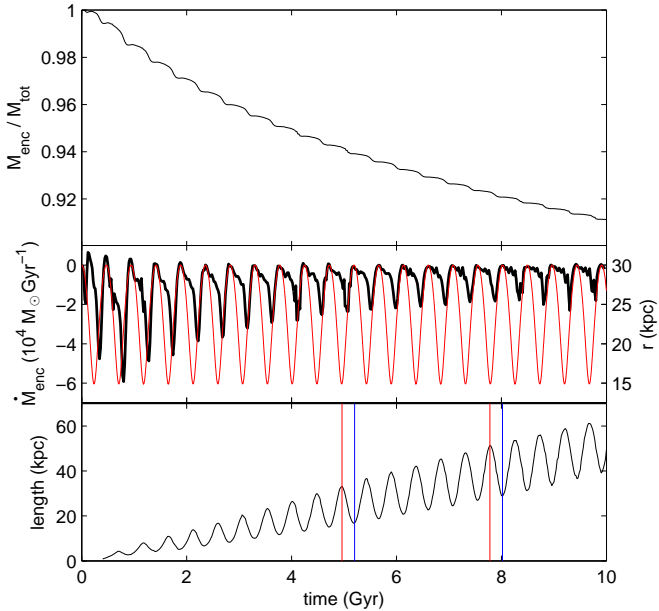}
	\caption{Top: fraction of mass enclosed in 0.103 kpc of the star cluster while orbiting the dark matter
	halo without any subhalos. Middle: rate of change of mass enclosed in 0.103 kpc of the star cluster (solid black)
	and radial position of the star cluster's orbit (red). This shows that the star cluster experiences bursts
	of mass loss almost immediately after each pericentric passage. \newrevision{Bottom: length of the stream. This shows that
	the stream gets stretched and compressed depending on its orbital phase. The red and blue vertical lines correspond to
	the 	snapshots where the star cluster is at the pericenters and apocenters of its orbit, respectively. These times
	are selected to demonstrate the bias discussed in Section \ref{sec:case_study}.}}
	
%	\caption{Top: Fraction of mass enclosed in 0.103 kpc of the star cluster while orbiting the dark matter
%	halo without any subhalos. Middle: Rate of mass loss enclosed in 0.103 kpc.
%	Bottom: Radial position of the star cluster's orbit. The mass loss is most severe right after each
%	pericentric passage since the mass loss is driven by the varying tidal field as the cluster oscillates
%	radially.}
	\label{fig:massloss}
\end{figure}

The star cluster is modeled as an N-body system which forms a stream as the cluster is disrupted by the tidal
field of the massive host. When the cluster is isolated, the energies of the individual particles
are conserved to a few percent over 10 Gyr. Using the softening as minimal impact distance, the relaxation
timescale in the core is $\gtrsim 110\, \mathrm{Gyr}$, which is much greater
than the orbital period at $\lesssim1\,\mathrm{Gyr}$. \fig{massloss} shows the mass enclosed inside 0.103 kpc of the
\newrevision{star cluster's} center \newrevision{it is orbiting in the absence of subhalos}. Because the stream is repeatedly stretched and compressed longitudinally along the
eccentric orbit, the mass enclosed in a fixed radius is not always decreasing in time. The bottom panel of 
\fig{massloss} shows that the mass loss is driven purely by bulge shocking \citep{bt08}, as the periodic bursts have exactly the
same period as the radial period of the orbit.

\begin{figure}
    \centering
    \includegraphics[width=3.5in]{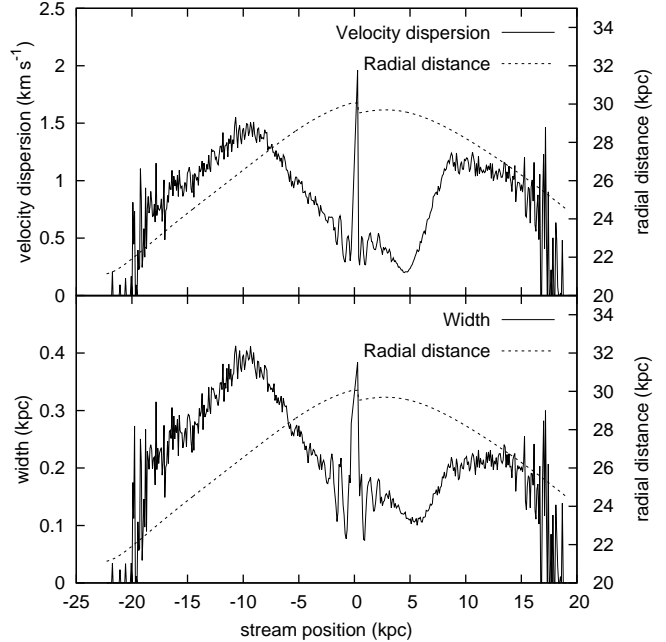}
    \caption{Tangential velocity dispersion and FWHM width along the stream at 8 Gyr. The progenitor is centered at 0 kpc, where
    positive and negative positions represent the leading and trailing branches of the stream, respectively.
    The structures in velocity dispersion and width are due to the eccentricity of the stream, as traced by
    the orbital distance along the stream.}
    \label{fig:stream_sigma_width}
\end{figure}

\fig{stream_sigma_width} shows the velocity dispersion and traverse FWHM of our simulated stream
without any subhalos. The stream is chosen
so that its properties are on the same orders of magnitude as Pal-5 \citep{dehnen04, odenkirchen09} and
GD-1 \citep{koposov10, willett09}. In the derivation in \citet{carlberg12}, the gap formation rate is expressed
as a function of galactocentric distance of the orbit and width of the stream. For our simulated stream, we adapt
average values of 22 kpc and 0.3 kpc, respectively, over the entire stream.

In a time independent and spherical potential $\Phi(r)$, three interesting conserved quantities are the radial,
azimuthal, and latitudinal actions where
\begin{eqnarray}
	J_r = \frac{1}{\pi}\int_{r_p}^{r_a} dr \sqrt{2E-2\Phi(r)-\frac{L^2}{r}} \\
	J_\phi = L_z \\
	J_\theta = L-|L_z|
\end{eqnarray}
respectively \citep{bt08}, where $r_a$ and $r_p$ are the apo- and pericentric distances of the orbit,
respectively. Since our stream's orbital plane is the $xy$-plane, $J_\theta\approx 0$ (though not exactly 0
because the stream has a finite thickness),
so we only consider $J_r$ and $L_z$. For simplicity, we ignore the progenitor and only consider the
particles which have already escaped from the cluster,
so when computing $J_r$ we assume that the potential due to the progenitor's potential is negligible.

\begin{figure}
	\centering
	\includegraphics[width=3.5in]{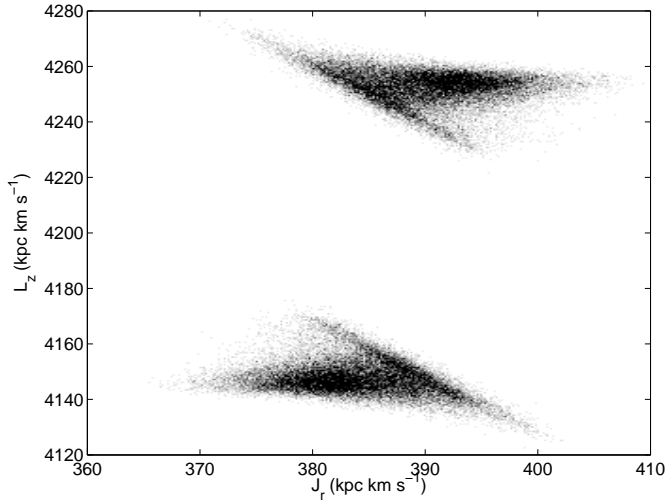}
	\caption{Distribution of orbital action variables $(J_r,L_z)$ for our stream at 8 Gyr without any subhalos. Along the entire
	stream with the progenitor masked out, 50,000 unbound particles were randomly selected to be placed on this map. Each pixel in this
	map represents the density of particles in that bin. The two symmetric lobes are the two branches of the stream.
%	Note that the relative spread in $J_r$ is about 10 times
%	larger than in $L_z$, so the spread in $J_r$ dominates the evolution of any features that traverse the width
%	of a stream. 
	Similar plots can be found in \citet{EB11}, and a plot in scaled energy and angular momentum
	in \citet{yoon11}.}
	\label{fig:Jr_Lz_distribution}
\end{figure}

$J_r$ and $L_z$ are useful since their dispersions are the origins of the stream's average width.
For example, in the epicyclic approximation \citep{bt08} 
where $\kappa$ and $a$ are the epicyclic frequency and amplitude, respectively, the radial motion can be written as
\begin{equation}
	r(t)=a \cos(\kappa t + \psi)
\end{equation}
where $\psi$ is an arbitrary phase angle. Then it can be shown that
\begin{eqnarray}
	J_r = \frac{1}{2\pi} \oint p_r dr \propto \int \dot{r}\,dr \propto \kappa a^2 \\
	L_z = R_g^2\Omega
\end{eqnarray}
where $\Omega$ and $R_g$ are the orbital frequency and radius of the guiding center, respectively. Clearly, dispersions
in both $a$ and $R_g$ can affect the width of the stream. Therefore, conserved quantities $J_r$ and $L_z$
are especially valuable in understanding the width of the stream.
\fig{Jr_Lz_distribution} shows the distributions in $(J_r,L_z)$ when
the stream is 8 Gyr old. The two lobes at higher and lower $L_z$ are the trailing and leading branches of
the stream, respectively.
%The two diagonal stripes on the edge of each lobe maybe due to the progenitor's
%potential which is ignored, as those stripes contain particles that are very close to the progenitor
%along the stream.
The absolute dispersions in $J_r$ and $L_z$ are on the same order of magnitude, in rough agreement
with the formula $\Delta J_r/\Delta L\sim (r_a-r_p)/\pi r_p$ \citep{EB11}.
In Section \ref{sec:gap_morphology}, we will show how the spread in actions affects the morphology of stream gaps.

%However, if we compare the dispersions
%relative to the orbital values, then $\Delta J_r/J_r$ is about 10 times larger than $\Delta L_z/L_z$.
%This means that the spread in radial action is mostly responsible for the width of the stream.

Plots similar to \fig{Jr_Lz_distribution} can be found in \citet{EB11} where they used angle-action
variables extensively to study the relation between the stream and the orbit.
A similar plot can also be found in \citet{yoon11}, but in scaled energy and
angular momentum which were first used by \citet{johnston98} to describe the dynamics of tidal streams.

%, but the energy is an abstract quantity 
%which is more difficult to interpret.

% --------------- Jr Lz distribution ---------------

% --------------------------------------------------

\section{Method}

\subsection{Subhalo Mass Ranges}
\label{sec:subhalo_mass_ranges}

%We simulate the stream fourteen times with identical initial conditions for the satellite, but with different subhalos.
\citet{yoon11} divided their subhalo mass spectrum into separate mass ranges in order to resolve the contributions
from each mass range. Using the same approach, we divide the subhalos
from $6.5\sci{4}$ to $10^8\Msun$ into 13 mass ranges (\tab{subhalo_massrange}). Each mass range contains incrementally
more subhalos, starting from the higher mass end toward the lower mass end. In the higher mass end, the mass ranges are chosen
such that the increase in subhalo masses are roughly the same. In the lower mass end, the mass ranges are chosen such that the
increase in subhalo numbers are roughly the same.

In each set of subhalos, we reduce their numbers by eliminating those whose orbits are always inside the perigalacticon
and outside the apogalacticon of the \newrevision{progenitor's orbit}. The largest subhalo in our simulations has a length scale of $\sim 1\,\mathrm{kpc}$
(\eqn{subhalo_size}), so \newrevision{all the subhalos with perigalacticon (apogalacticon) larger (smaller) than that of
the progenitor's orbit by $\sim2\,\mathrm{kpc}$ will interact minimally with the stream. This allows us to safely eliminate
the subhalos with perigalacticon larger than $32\,\mathrm{kpc}$, and apogalacticon smaller than $13\,\mathrm{kpc}$.}

%In each set of subhalos, we reduce their numbers by eliminating those whose orbits are always inside the perigalacticon
%and outside the apogalacticon of the smooth stream. The largest subhalo in our simulations has length scale of $\sim 1$ kpc
%(\eqn{subhalo_size}), so we can safely eliminate those subhalos with perigalacticon larger than 32 kpc, and apogalacticon
%smaller than 13 kpc.

We run 14 simulations -- 1 ``smooth stream'' without any subhalos, and 13 ``\lcdm\ streams'' containing
the subhalos in the mass ranges in \tab{subhalo_massrange} -- with identical initial conditions and dark matter halo
potential. This allows us to resolve the effects of the lower mass subhalos whose existence are in question.

%Each individual range
%contains roughly the same number of subhalos (in the lower mass end) or roughly in the same total mass $\sim1.5\sci{10}\Msun$
%in subhalos (for the higher mass end). We run fourteen simulations --
%one ``smooth stream'' without any subhalos, and thirteen ``\lcdm\ streams'' containing the
%subhalos in the mass ranges summarized in \tab{subhalo_massrange} -- with identical initial conditions and dark matter halo
%potential. Each \lcdm\ stream simulation contains incrementally more subhalos starting from the higher mass end towards the
%lower mass end. This allows us to resolve the effects of the lower mass subhalos whose existence are in question.

\begin{deluxetable}{cccccc}
    	\tablecaption{Thirteen Subhalo Mass Ranges in Our Simulations.}
    \tablehead{
    \colhead{$m_{low}/\Msun$} & \colhead{$N$} & \colhead{$M/\Msun$} & \colhead{$N_{orbit}$} & \colhead{$M_{orbit}/\Msun$}}
	\startdata
    $5.3\sci{7}$ & 203     & $1.5\sci{10}$  & 30     & $2.2\sci{9}$ \\
    $2.7\sci{7}$ & 593     & $2.9\sci{10}$  & 98     & $4.7\sci{9}$ \\
    $1.3\sci{7}$ & 1,392   & $4.4\sci{10}$  & 220    & $6.9\sci{9}$ \\
    $5.8\sci{6}$ & 3,160   & $5.9\sci{10}$  & 476    & $9.1\sci{9}$ \\
    $2.5\sci{6}$ & 7,038   & $7.3\sci{10}$  & 1,101  & $1.1\sci{10}$ \\
    $1.0\sci{6}$ & 16,394  & $8.8\sci{10}$  & 2,576  & $1.4\sci{10}$ \\
    $3.6\sci{5}$ & 41,515  & $1.0\sci{11}$  & 6,539  & $1.6\sci{10}$ \\
    $2.1\sci{5}$ & 67,599  & $1.1\sci{11}$  & 10,563 & $1.7\sci{10}$ \\
    $1.5\sci{5}$ & 91,601  & $1.1\sci{11}$  & 14,337 & $1.8\sci{10}$ \\
    $1.1\sci{5}$ & 121,181 & $1.2\sci{11}$  & 18,872 & $1.8\sci{10}$ \\
    $9.0\sci{4}$ & 145,220 & $1.2\sci{11}$  & 22,578 & $1.9\sci{10}$ \\
	$7.5\sci{4}$ & 171,163 & $1.2\sci{11}$  & 26,586 & $1.9\sci{10}$ \\
	$6.5\sci{4}$ & 194,726 & $1.2\sci{11}$  & 30,253 & $1.9\sci{10}$

%    $6.5\sci{4}$ & $7.5\sci{4}$ & 23563 & $1.6\sci{9}$  & 3667 & $2.6\sci{8}$ \\
%    $7.5\sci{4}$ & $9.0\sci{4}$ & 25942 & $2.1\sci{9}$  & 4008 & $3.3\sci{8}$ \\
%    $9.0\sci{4}$ & $1.1\sci{5}$ & 24039 & $2.4\sci{9}$  & 3706 & $3.7\sci{8}$ \\
%    $1.1\sci{5}$ & $1.5\sci{5}$ & 29580 & $3.8\sci{9}$  & 4535 & $5.8\sci{8}$ \\
%    $1.5\sci{5}$ & $2.1\sci{5}$ & 24002 & $4.2\sci{9}$  & 3774 & $6.7\sci{8}$ \\
%    $2.1\sci{5}$ & $3.6\sci{5}$ & 26084 & $7.1\sci{9}$  & 4024 & $1.1\sci{9}$ \\
%    $3.6\sci{5}$ & $1.0\sci{6}$ & 25121 & $1.5\sci{10}$ & 3963 & $2.3\sci{9}$ \\
%    $1.0\sci{6}$ & $2.5\sci{6}$ & 9355  & $1.4\sci{10}$ & 1475 & $2.3\sci{9}$ \\
%    $2.5\sci{6}$ & $5.8\sci{6}$ & 3878  & $1.4\sci{10}$ & 625  & $2.3\sci{9}$ \\
%    $5.8\sci{6}$ & $1.3\sci{7}$ & 1768  & $1.5\sci{10}$ & 256  & $2.2\sci{9}$ \\
%    $1.3\sci{7}$ & $2.7\sci{7}$ & 798   & $1.5\sci{10}$ & 122  & $2.2\sci{9}$ \\
%    $2.7\sci{7}$ & $5.3\sci{7}$ & 390   & $1.5\sci{10}$ & 68   & $2.5\sci{9}$ \\
%    $5.3\sci{7}$ & $1.0\sci{8}$ & 203   & $1.5\sci{10}$ & 30   & $2.2\sci{9}$ 
	\enddata
	\tablenotetext{}{The upper limit cuts off at $10^8\Msun$
	for all ranges. Columns from left to right: lower mass limit,
	total number and total mass in subhalos (\eqn{subhalo_massfunction}), number and mass in subhalos used in simulation
	after reduction by orbit (see the text). \label{table:subhalo_massrange}}
\end{deluxetable}

\subsection{Gap Finding}
\label{sec:gapfinding}

Gaps are manifested as local minima in the linear density along the stream. To obtain the linear
density along a stream in an eccentric orbit, we first fit the stream with two degree-6 polynomials --
one each for the leading and trailing streams -- in polar coordinates centered at the Galactic center. The 
points along each line are spaced at 0.002 radians apart. Between each pair of adjacent points a cylinder
of radius 1 kpc is drawn which lies lengthwise along the pair of points. The linear density is then the number of particles
inside this cylinder divided by the length of the cylinder. This spacing is chosen so that the gaps
as wide as the stream are well resolved.

\begin{figure}
    \centering
    \includegraphics[width=3.5in]{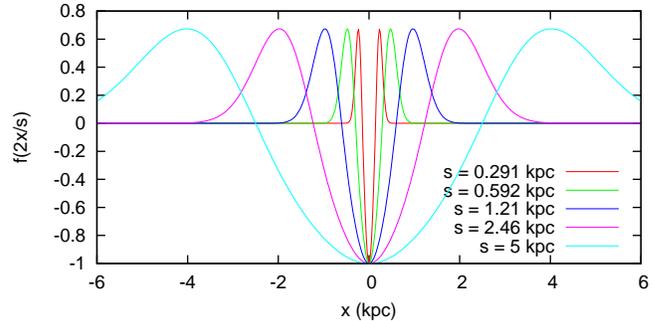}
    \caption{Five example filter scales for the match filter used in \citet{pal5_gaps}, which has the functional form
    $f(t)=(t^6-1)\exp(-1.2321t^2)$. With a physical scale $s$, the integral of $f(2x/s)$ vanishes inside $-1.5s<x<1.5s$.
    The roots of each filter are located at $\pm s/2$ so that the gap length is simply the distance
    between the roots. We search for gaps at 12 logarithmically filter scales from 0.1 to $5\,\mathrm{kpc}$
    in order to minimize the chance of detecting the same gap at multiple scales.}
    \label{fig:filter_scales}
\end{figure}

%A subhalo gap is manifested with a characteristic shape -- a local minimum in linear density with two local maxima on both
%sides of the minimum due to conservation of mass -- along the stream. 
The method used to find gaps in stream densities is inspired by
the technique first used by \citet{pal5_gaps} to find gaps in observations. They used matched filters of the estimated
shape of a density gap at various length scales to look for positions in the stream which potentially contain gap signals.
The filter consists of a local minimum which is the underdensity of stars, and two local maxima on both sides of the
minimum due to conservation of mass (\fig{filter_scales}).
This method is similar to the wavelets approach, where the integral of the filter function is constructed to vanish
inside a certain domain. The potential gap signals are then easily identified as local maxima in the convolution between
the filter and the signal.

To obtain the significance of each potential gap signal against noise, \citet{pal5_gaps} produced bootstrap samples from
the sky background. With the simulations in this study,
we can estimate noise levels using the smooth stream. Note that the ``smooth stream'' 
itself is not totally smooth. As we will show in Section \ref{sec:gap_counting}, there are large density fluctuations near
the progenitor due to the coherent in epicyclic motion
of the particles, as first explained by \citet{kupper08}. When the particles become unbound from the progenitor, they pile up
near the base of their cycloid trajectories, creating epicyclic overdensities (hereafter EO) along the stream.
Although this intrinsic process to mimic gaps can be confused with gaps caused by subhalos, EOs are only
\newrevision{apparent within $\lesssim5\,\mathrm{kpc}$ away from the progenitor in our streams
(\figs{density_gaps_whole_massrange0}{density_gaps_central_massrange0}). The details of the dynamics of EOs is beyond
the scope of this study, but this effect can be understood in terms of orbital actions. EOs occur due to coherent
epicyclic motions of the particles, which nevertheless have finite dispersions in orbital actions (\fig{Jr_Lz_distribution})
and are not perfectly coherent. Therefore, although the escaping particles' orbits stay roughly coherent in the first few clumps,
their orbits eventually drift out of phase as they travel along the stream. This explains why the density peaks of EOs
further downstream are not as apparent as the peaks closer to the progenitor \citep{just09, kupper10}.}

After masking $10\,\mathrm{kpc}$ of the smooth stream centered at the progenitor, the rest of the
smooth stream is simply noise. Our method to find gaps in a given stream can be summarized as follows.

\begin{enumerate}
%    \item Fit a smooth baseline $\bar\rho(x)$ against the overall shape of the stream density $\rho(x)$. The baseline is then subtracted from
%    the density to obtain a density difference.
    
    \item Compute
    \begin{equation}
    		C_s(x) = \frac{1}{s} \int_{x-1.5s}^{x+1.5s} \left[ \rho(x^\prime)-\bar\rho_s(x) \right]\, f\left( \frac{2(x-x^\prime)}{s} \right) dx^\prime
		\label{eqn:cs}
%    		C_s(x) = \frac{1}{s} \int_{-\infty}^\infty \left[\rho(x^\prime)-\bar\rho(x^\prime) \right]\, f\left( \frac{2(x-x^\prime)}{s} \right) dx^\prime
    \end{equation}
    where $f(t)=(t^6-1)\exp(-1.2321t^2)$ is a matched filter function \citep{pal5_gaps,CG13},
    and $s$ is the filter scale. $\bar\rho_s(x)$ is the mean of $\rho(x)$ inside $[x-1.5s,x+1.5s]$,
    the domain in which the integral of $f(2x/s)$ itself vanishes.
    Each potential gap signal would appear as a local
    maximum in $C_s(x)$. This convolution is computed at 12 logarithmically spaced filter scales
    from 0.1 to $5\,\mathrm{kpc}$ (\fig{filter_scales}), and then all the local
    maxima of each stream are sorted by $C_s$.
    
    \item Repeat the above step using the smooth stream, but with $\pm 5$ kpc from the progenitor masked along the stream.
    The set of local maxima in $\tilde C_s(x)$ from this convolution is the noise, which are also sorted by $\tilde C_s$.
    
    \item
    Each local maximum in the signal set are compared against the noise set. A signal element at any position along
    either branch of the stream that ranks higher than 99\% in the noise set is identified as a gap.
    
\end{enumerate}

Inevitably, this method may detect the same gap at 99\% confidence at different scales but in very close proximity.
To avoid over-counting, we employ the following scheme to eliminate overlapping gaps. First, we define an overlap as
two gaps whose $C_s$ local maxima are identified at $C_{s1}(x_1)$ and $C_{s2}(x_2)$ that are within $s_1$ away from
each other along the stream, where $s_2<s_1$. When this occurs, the gap with higher $C_s$ eliminates the lower.

\newrevision{Our gap detection method requires no prior knowledge whether a given gap is an EO or a subhalo perturbation, both of which
can be identified as a series of over- and under-densities. When we count the
number of gaps in the end, EOs will be included. One key result of our study is that gaps due to EOs
are distributed very differently in lengths compared to gaps due to subhalo perturbation.}

\section{Results and discussion}

\subsection{Gap Morphology}
\label{sec:gap_morphology}

According to \citet{yoon11}, gaps in general are diagonal and not perpendicular to the stream due to a gradient
in angular momentum (hence a gradient in orbital velocities) across the width of the stream, which can shear a gap longitudinally.
\fig{Jr_Lz_distribution} allows us to estimate the shearing effect in our streams using the distributions in
angular momenta. For each branch of the stream, the FWHM spread in angular momentum is about
$\Delta L \sim \Delta L_z \sim10\,\mathrm{kpc}\,\mathkms$. For a narrow stream at $r=22\,\mathrm{kpc}$, the spread
in velocity is $\Delta v = \Delta L/r \sim 0.5\,\mathkms$. Therefore, a gap that spans the width of the
stream will be sheared by less than $1\,\mathrm{kpc}$ per Gyr.

\begin{figure}
	\centering
	\includegraphics[width=3.5in]{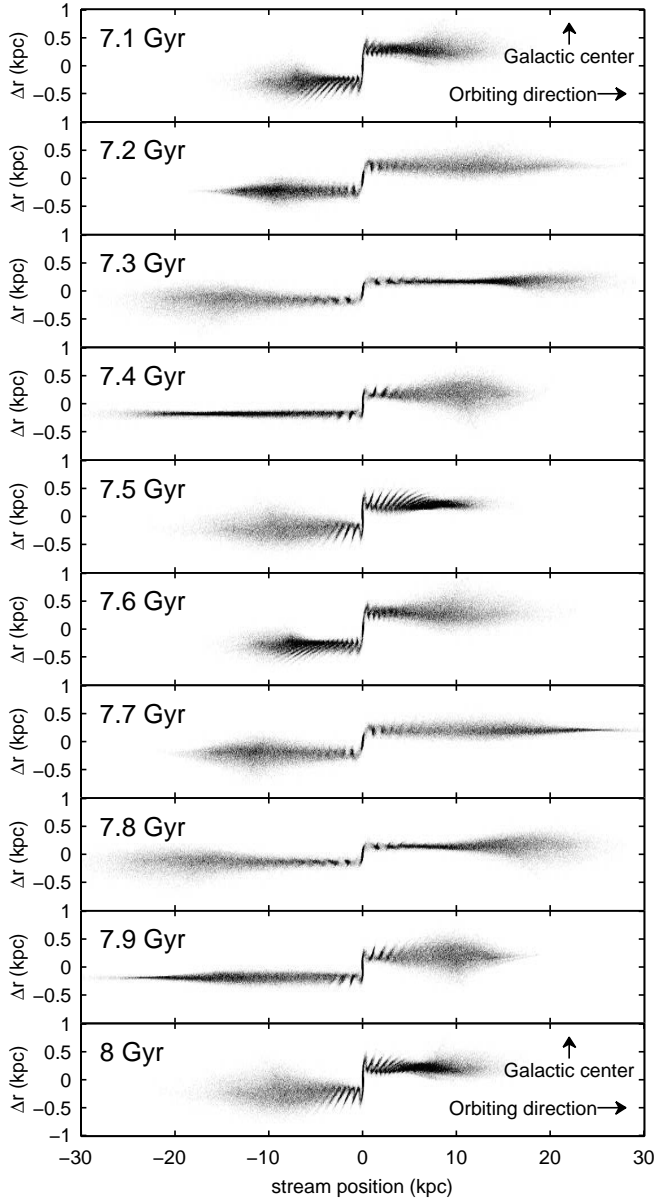}
	\caption{\newrevision{Surface density of the smooth stream from 7 to 8 Gyr projected onto the $xy$-plane. The stream is then aligned
	to Cartesian coordinates} where the horizontal axis is the offset
	position along the stream from the progenitor, and the vertical axis is the radial offset from the galactocentric
	distance of the progenitor. \newrevision{This is done by tracing a best fit line along the stream.} For each segment in the line,
	the particles in between the end points of the segment are rotated such that the Galactic center points toward the $+y$
	direction in this plot. Note that the vertical axis has been scaled 30 times the larger than the horizontal axis.}
	\label{fig:stream_maps_smooth}
\end{figure}

\begin{figure}
	\centering
	\includegraphics[width=3.5in]{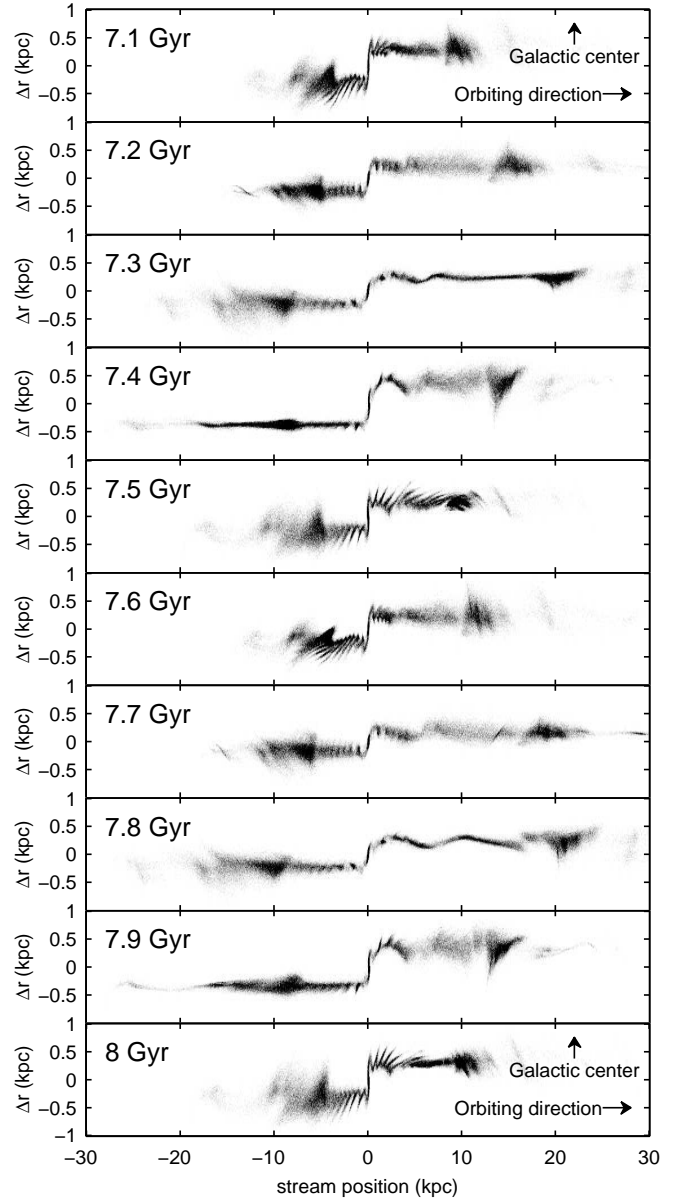}
	\caption{Stream with $5.8\sci{6}<M/\Msun<10^8$ subhalos from 7 to $8\,\mathrm{Gyr}$, \newrevision{aligned to} Cartesian coordinates
	similar to the smooth stream in \fig{stream_maps_smooth}. Compared to the smooth stream, this \lcdm\ stream
	shows much more structures at various scales. In general, whether by EOs
	(mostly inside $\pm5\,\mathrm{kpc}$) or by subhalo perturbations, gaps have complicated morphologies and do not even
	have straight edges across the width	of the stream.}
	\label{fig:stream_maps_lcdm}
\end{figure}

\figs{stream_maps_smooth}{stream_maps_lcdm} show the time evolution of the
smooth and a \lcdm\ stream, respectively, from 7 to $8\,\mathrm{Gyr}$. The EOs near the progenitor
appear to shear by different amounts at different times, but this is due to the radial oscillation in the orbit where the
radial period is $\sim0.5\,\mathrm{Gyr}$. Upon closer inspection of \fig{stream_maps_lcdm}, we also note
that not only do subhalo gaps have complicated morphologies, but their orientations flip back and forth in a radial period
due to the spread in $J_r$. Nevertheless, comparing panels of the same radial phase at one radial period apart, 
the end points of each gap across the width of the stream do not shift by any appreciable amount. Rather,
the morphologies of the subhalo gaps are already apparent as each gap first appears.

%Note that
%this gap is already slanted due to the the subhalo's impact angle on the stream. As the stream oscillates radially
%in its orbit, due to its spread in $J_r$ the orientation of the diagonal gap flips back and forth, but the two end
%points of the gap across the width of the stream did not shift by more than a few kpc.

If the linear density of a stream is calculated by integrating the entire thickness of the stream in traversing slices
along the stream, then the contrast of the gap will be reduced. This is because the edges of the gaps are not
perfectly straight across the width of the stream, so dividing the stream into slices will smear out the density
contrast. To investigate how much the smearing will affect gap detection, we 
calculate the linear densities in two ways. (1) Integrating cylindrical slices of radius 1 kpc along the stream,
hereafter the ``whole width,'' where 1 kpc was chosen to cover the entire thickness of the whole stream.
(2) Integrating only the cylindrical slices of radius 0.04 kpc centered along the best fit line of 
each branch of the stream, hereafter the ``central width,'' where 0.04 kpc is chosen to mimic the GD-1's observed width
of 0.08 kpc \citep{CG13}. This central width then encloses about 30\%--40\% of the mass of the whole width, depending
on its orbital phase where, for example, the stream is radially compressed during pericentric passage.

% -------------- diagonal gaps -------------------

% -------------------------------------------------

\subsection{Gap Counting}
\label{sec:gap_counting}

\begin{figure}
	\centering
	\includegraphics[width=3.5in]{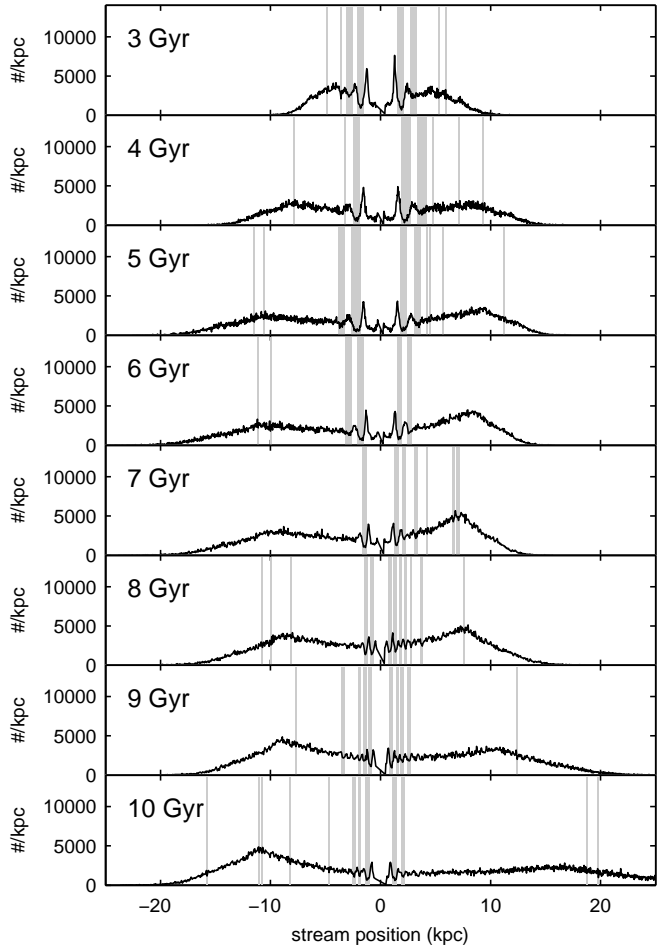}
	\caption{Linear densities along the smooth stream, integrated for the whole thickness of the stream, from 3 to $8\,\mathrm{Gyr}$.
	The progenitor is centered at $0\,\mathrm{kpc}$ and is masked out. Shaded columns are gaps identified at 99\% confidence at the
	scale depicted by the columns' widths.}
	\label{fig:density_gaps_whole_massrange0}
\end{figure}

\figsto{density_gaps_whole_massrange0}{density_gaps_central_massrange13} show the densities along
the whole and central widths of the smooth stream and a \lcdm\ stream from 3 to $10\,\mathrm{Gyr}$.
The streams younger than $3\,\mathrm{Gyr}$ are not shown as the stream is $\lesssim 10\,\mathrm{kpc}$ long
at those ages, so the gaps are dominated by very prominent EOs. Moreover, the stream itself does not yet have a large enough cross
section to produce enough gaps for meaningful statistics. In each panel, the shaded
columns represent the gaps that are found on the scale of the columns' widths. Although
these gaps are identified as being 99\% significant, the density contrasts of the gaps have not been
quantified. \newrevision{For the rest of this paper, we assume that all gaps identified at 99\% significance
can be observed.}
Note that because the gap finding process is applied independently to each snapshot, the shaded columns
do not necessarily represent the time evolutions of individual gaps. Instead, the shaded columns show the general
distributions of gaps -- both in space and in gap lengths.

\begin{figure}
	\centering
	\includegraphics[width=3.5in]{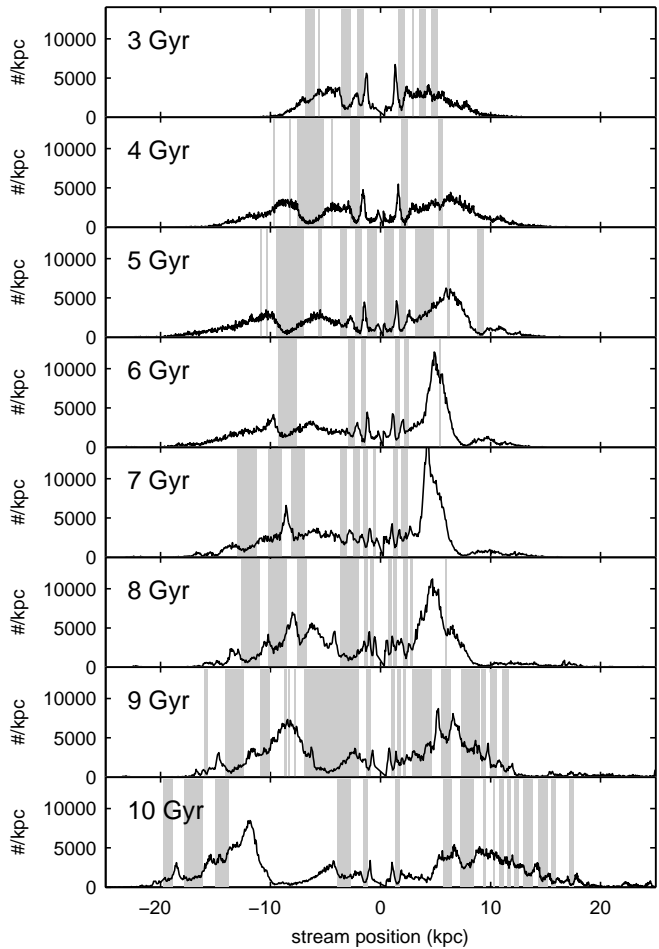}
	\caption{Linear densities and gaps identified at 99\% in the \lcdm\ stream with subhalo masses $6.5\sci{4}<M/\Msun<10^8$, integrated for the 
	whole thickness of the stream.}
	\label{fig:density_gaps_whole_massrange13}
\end{figure}

Our gap finding method is has a number of problems. In the smooth stream
(\figs{density_gaps_whole_massrange0}{density_gaps_central_massrange0}), our method by construction
identifies 1\% of the noise as gaps. This is why there can be spurious gaps detected well beyond 5 kpc away from the
progenitor, even though EOs tend to form very close to the progenitor. Also, the overall profile
of the stream density can sometimes be confused as a gap as well. One example is a gap at $4\,\mathrm{kpc}$ at $5\,\mathrm{Gyr}$
in \fig{density_gaps_whole_massrange13},
where a smooth density gradient from 3 to $8\,\mathrm{kpc}$ is mistaken as the right half of a long gap. At the 95\%
confidence limit, both kinds of false positives are quite common and can often be identified by eye. For the results
below, we show gaps that are 99\% significant, which minimizes the occurrence false positives.

% --------------- densities -----------------------

\begin{figure}
	\centering
	\includegraphics[width=3.5in]{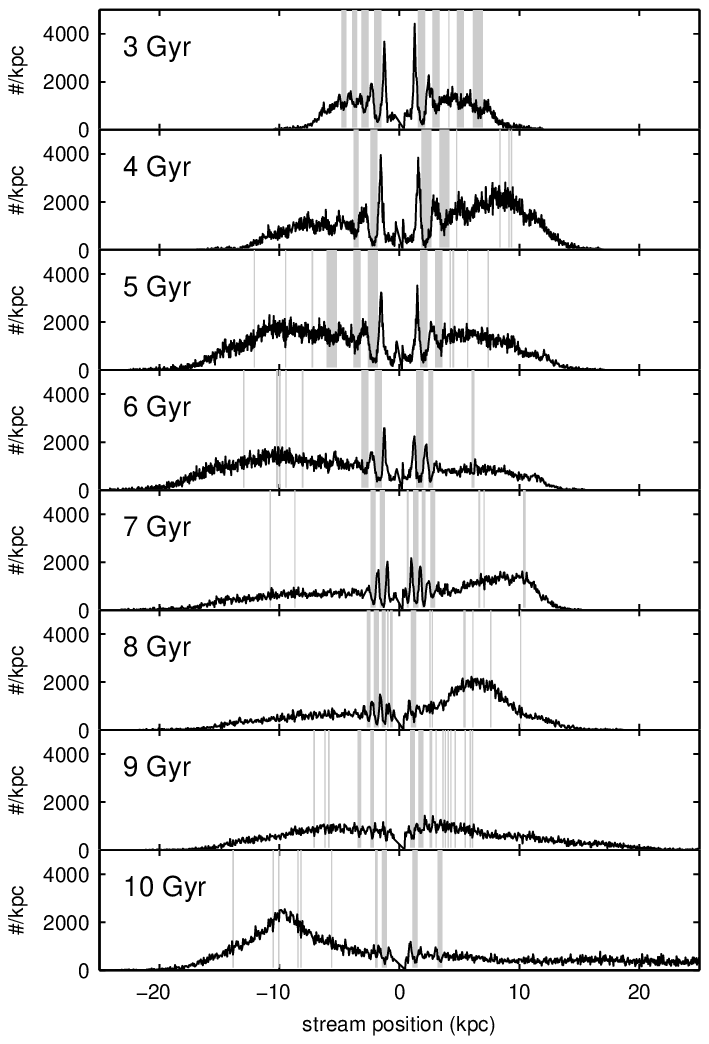}
	\caption{Linear densities and gaps identified at 99\% in the smooth stream, but integrated for only
	a cylinder of diameter $0.08\,\mathrm{kpc}$ along the central line of the stream.}
	\label{fig:density_gaps_central_massrange0}
\end{figure}

%\begin{figure}
%	\centering
%	\includegraphics[width=3.5in]{figs/gap_finding/king12_1M_halo30_v1.5_mass5.3e7_1e8_1_gaps_cl0.99.eps}
%	\caption{Linear densities and gaps identified at 99\% in the \lcdm\ stream with subhalo masses $5.3\sci{7}<M/\Msun<10^8$, integrated for the 
%	whole thickness of the stream.}
%	\label{fig:density_gaps_whole_massrange1}
%\end{figure}
%\begin{figure}
%	\centering
%	\includegraphics[width=3.5in]{figs/gap_finding/king12_1M_halo30_v1.5_mass5.3e7_1e8_1_central0.04_gaps_cl0.99.eps}
%	\caption{Linear densities and gaps identified at 99\% in the \lcdm\ stream with subhalo masses $5.3\sci{7}<M/\Msun<10^8$, integrated for only
%	a cylinder of diameter $0.08\,\mathrm{kpc}$ along the central line of the stream.}
%	\label{fig:density_gaps_central_massrange1}
%\end{figure}

\begin{figure}
	\centering
	\includegraphics[width=3.5in]{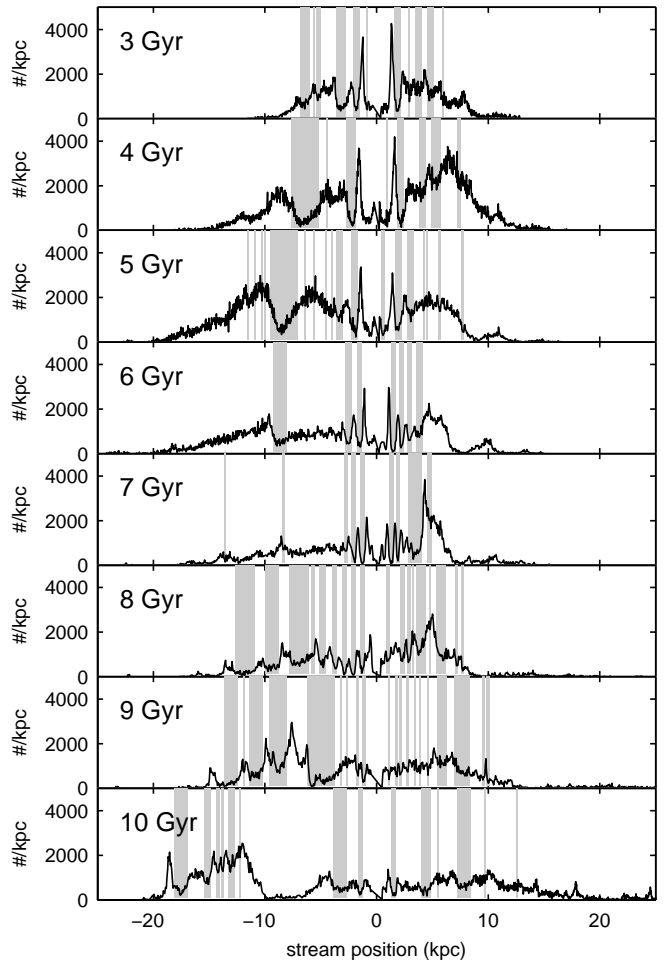}
	\caption{Linear densities and gaps identified at 99\% in the \lcdm\ stream with subhalo masses $6.5\sci{4}<M/\Msun<10^8$, integrate for only
	a cylinder of diameter $0.08\,\mathrm{kpc}$ along the central line of the stream.}
	\label{fig:density_gaps_central_massrange13}
\end{figure}

% ---------------------------------------------------------

%$\rho$ and smooth baseline $\bar\rho$,
%and the bottom panel is $\delta\equiv (\rho-\bar\rho)/\bar\rho$, with the gaps identified by shaded columns.

\subsection{Gap Spectrum}
\label{sec:gap_size_distribution}

%\figsto{density_gaps_whole_massrange0}{density_gaps_central_massrange13} show the locations and lengths of gaps
%identified in the whole and central widths of the smooth stream and a \lcdm\ stream. 
%The streams younger than $3\,\mathrm{Gyr}$ are not shown as the stream is $\lesssim 10\,\mathrm{kpc}$ long
%at those ages, so the gaps are dominated by very prominent EOs. Moreover, the stream itself does not yet have a large enough cross
%section to produce enough gaps for meaningful statistics.
%Note that because the gap detection process is applied independently to each snapshot, the shaded columns
%do not necessarily represent the time evolutions of individual gaps. Instead, the shaded columns show the general
%distributions of gaps -- both in space and in gap lengths.

Following the idealized experiments in \citet{carlberg12}, \citet{CG13} derived an analytical relation between gap formation rate
$R_\cup$ which is the cumulative number of gaps longer than length $l$ per unit stream length per unit time as a function
of gap length (hereafter the ``gap spectrum'') such that
\begin{equation}
    R_\cup^{\mathrm{ideal}} = 0.060\, \hat{r}^{0.44}\, l^{-1.16}\, \mathrm{kpc}^{-1}\,\mathrm{Gyr}^{-1}
    \label{eqn:gaprate_gapsize_formula}
\end{equation}
where $\hat{r}\equiv r/30\,\mathrm{kpc}$, and we adapt $r=22\,\mathrm{kpc}$ for the average galactocentric radius
of the stream. In this section we aim to study the validity of \eqn{gaprate_gapsize_formula} in our
self-consistent stream. We set the ``length'' of each gap as the scale $s$ of the matched filter which identified
the gap (Section \ref{sec:gapfinding})

\subsubsection{Smooth Stream without Subhalos}

\figs{density_gaps_whole_massrange0}{density_gaps_central_massrange0} show the gaps identified in the smooth stream integrated using
the whole thickness and central thickness, respectively. Clearly, the gaps due to EOs are clustered at $\lesssim5\,\mathrm{kpc}$ 
on both sides of the progenitor, and all the gaps have very similar sizes. The measured $R_\cup$ would peak at short gaps and
quickly drops off beyond $l\gtrsim 1\,\mathrm{kpc}$. \eqn{gaprate_gapsize_formula} is meant to describe an idealized gap
spectrum produced by subhalos, and not by EOs.

A key result of this study, as discussed below, is that the gap spectrum for subhalo gaps looks very different than the gap spectrum
for EOs. The existence of gaps longer than $\sim 1\,\mathrm{kpc}$ would be an indication that
processes other than EOs are responsible for the gaps. Furthermore, subhalo gaps can be found everywhere along the
whole stream. EOs can only be observed in the immediate vicinity of the progenitor.

\subsubsection{\lcdm\ Stream with Independent Sets of Subhalos}

As an ideal case, \eqn{gaprate_gapsize_formula} ignores the visibility of gaps when the same position
of a stream suffers impacts by multiple subhalos at different times. For instance, after one major impact by 
a massive subhalo which results in a long and high contrast density gap at an early time, subsequent impacts by less massive subhalos
in that same region at a later time may not be visible. 

Gap overlapping can be minimized by the following experiment. We run 13 separate simulations with the same
initial conditions as the star cluster, but the subhalo masses are selected differentially from \tab{subhalo_massrange}.
This allows each stream to interact with an independent set of subhalos of a very \newrevision{small} range of masses. 
Overlapping can still occur within the same simulation for each set of subhalos (hence a small number of
gaps can still be eliminated), but to a much lesser extent than using integrated mass ranges.

\begin{figure}
	\centering
	\includegraphics[width=3.5in]{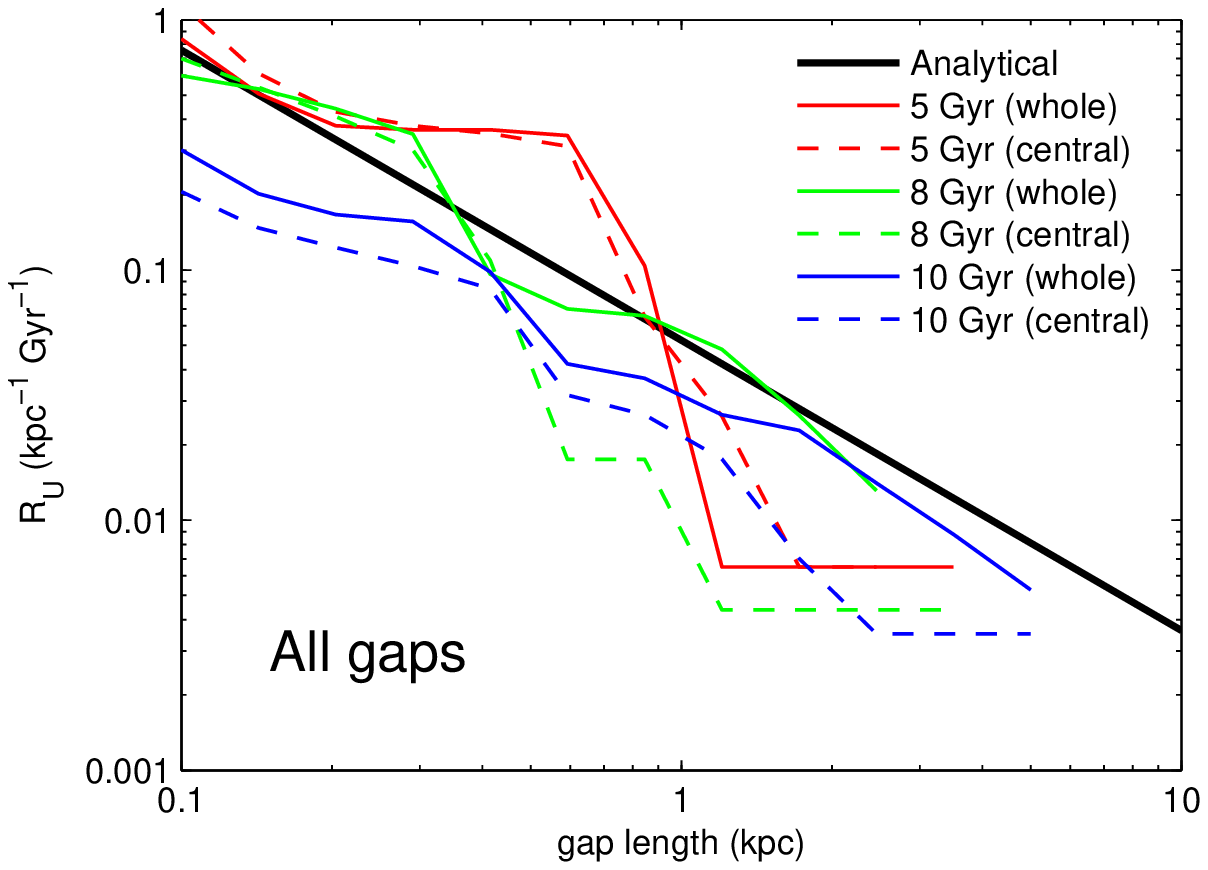}
	\includegraphics[width=3.5in]{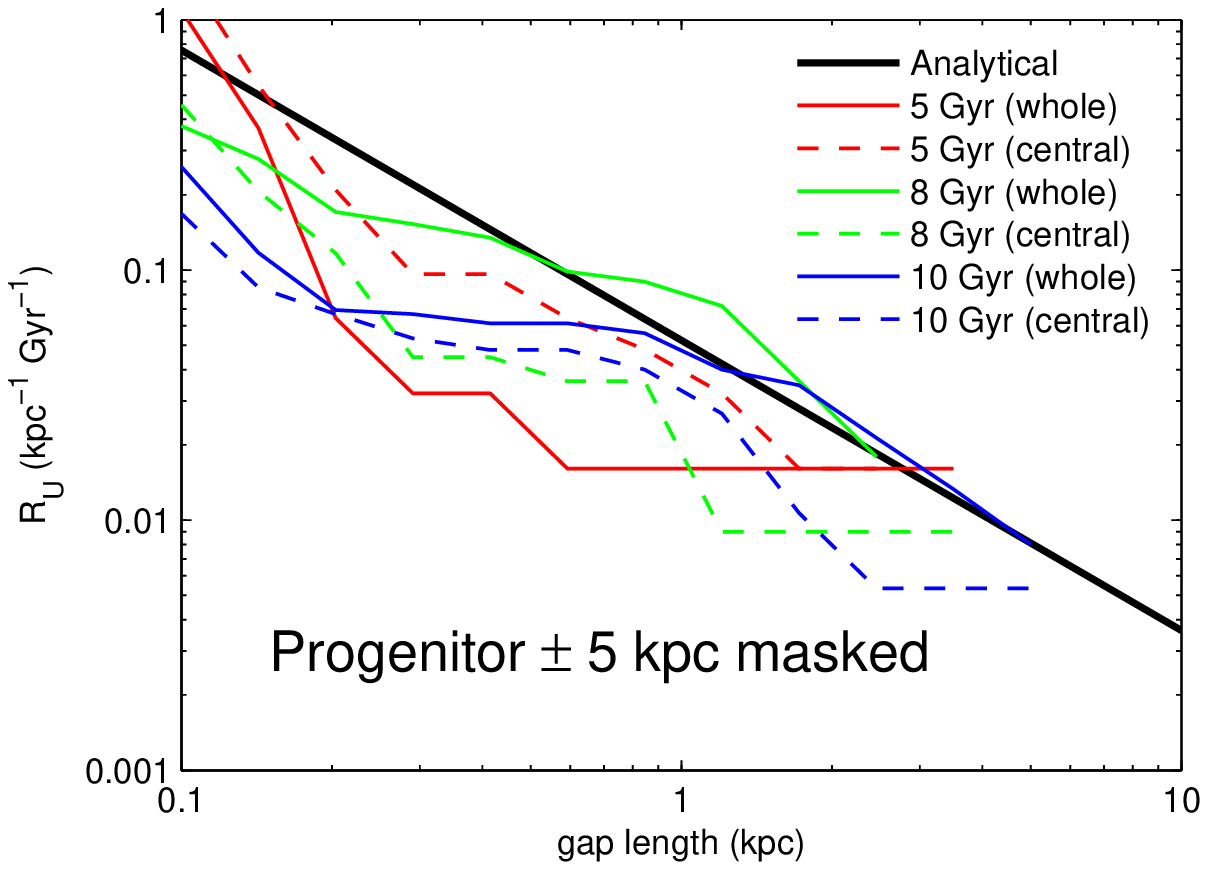}
	\caption{Analytical \citep{carlberg12} and simulated gap spectra. The gaps are collected from 13 simulations with independent
	sets of subhalos selected differentially from the mass ranges in \tab{subhalo_massrange}. In the top panel, all
	gaps identified (including EOs) \newrevision{are} included in the measured gap spectra. In the bottom panel, the gaps located within 
	$5\,\mathrm{kpc}$ away from the progenitor \newrevision{are} eliminated, and the stream lengths and ages \newrevision{are} adjusted.
	\newrevision{Since the gaps are mostly independent, the gap spectra follow the analytical prediction reasonably well.}
	}
	\label{fig:RU_summed}
\end{figure}

\fig{RU_summed} shows the measured gap spectrum from the gaps collected from all 13 simulations using independent sets of
subhalos. In the top
panel which includes all gaps, the measured gap spectrum matches \eqn{gaprate_gapsize_formula} reasonably well. However,
this is a coincidence as the gaps contain EOs which are not described by \eqn{gaprate_gapsize_formula}.
In an attempt to eliminate EOs, in the bottom panel of \fig{RU_summed}, the gaps that are located within $5\,\mathrm{kpc}$
away from the progenitor are eliminated. When computing the gap formation rates in these cases, the number of gaps are divided by
a stream length which is reduced by $10\,\mathrm{kpc}$ and a stream age which reduced by $2\,\mathrm{Gyr}$ (i.e.,
the age of the stream when it is $10\,\mathrm{kpc}$ long). This allows us to facilitate a fair comparison of gap
spectra against the cases which include all gaps in the entire stream.
Comparing the two panels in \fig{RU_summed}, we can see the masking of the $10\,\mathrm{kpc}$ around the progenitor
reduces the abundance of shorter gaps. This is expected since that region of the stream contains mostly EOs which occur
at scales $\lesssim 1\,\mathrm{kpc}$. In general it is difficult to tell whether a given gap within 
$5\,\mathrm{kpc}$ is due to EOs or subhalos, so in the process some subhalo gaps near the progenitor may have
been eliminated as well. 

Whether \eqn{gaprate_gapsize_formula} is a good description of the gap spectrum in a general stream likely requires
more simulations with varying orbital parameters. Nevertheless, it is apparent that the gap spectrum does not depend strongly on
the age or the integrating width for the linear density of the stream. However, it is still an ideal case since a stream
realistically interacts with all the subhalos at the same time. As we show in the following section, gap overlapping can
significantly alter the gap spectrum.

%In the top panel of \fig{RU_summed} where all gaps
%are included, and there is a clear excess at short gaps, which is due to EOs. In the bottom panel, the gaps within $5\,\mathrm{kpc}$
%from the progenitor are all eliminated. To facilitate a fair comparison of gap formation rate, since we are not
%considering the $10\,\mathrm{kpc}$ around the progenitor, the stream length is decreased by $10\,\mathrm{kpc}$, and the
%stream age is decreased by $2\,\mathrm{Gyr}$ which is the age of the stream when the its length is $10\,\mathrm{kpc}$. 
% 

\subsubsection{\lcdm\ Stream with all Subhalos}
\label{sec:lcdm_stream_with_all_subhalos}

\begin{figure}
	\centering
	\includegraphics[width=3.5in]{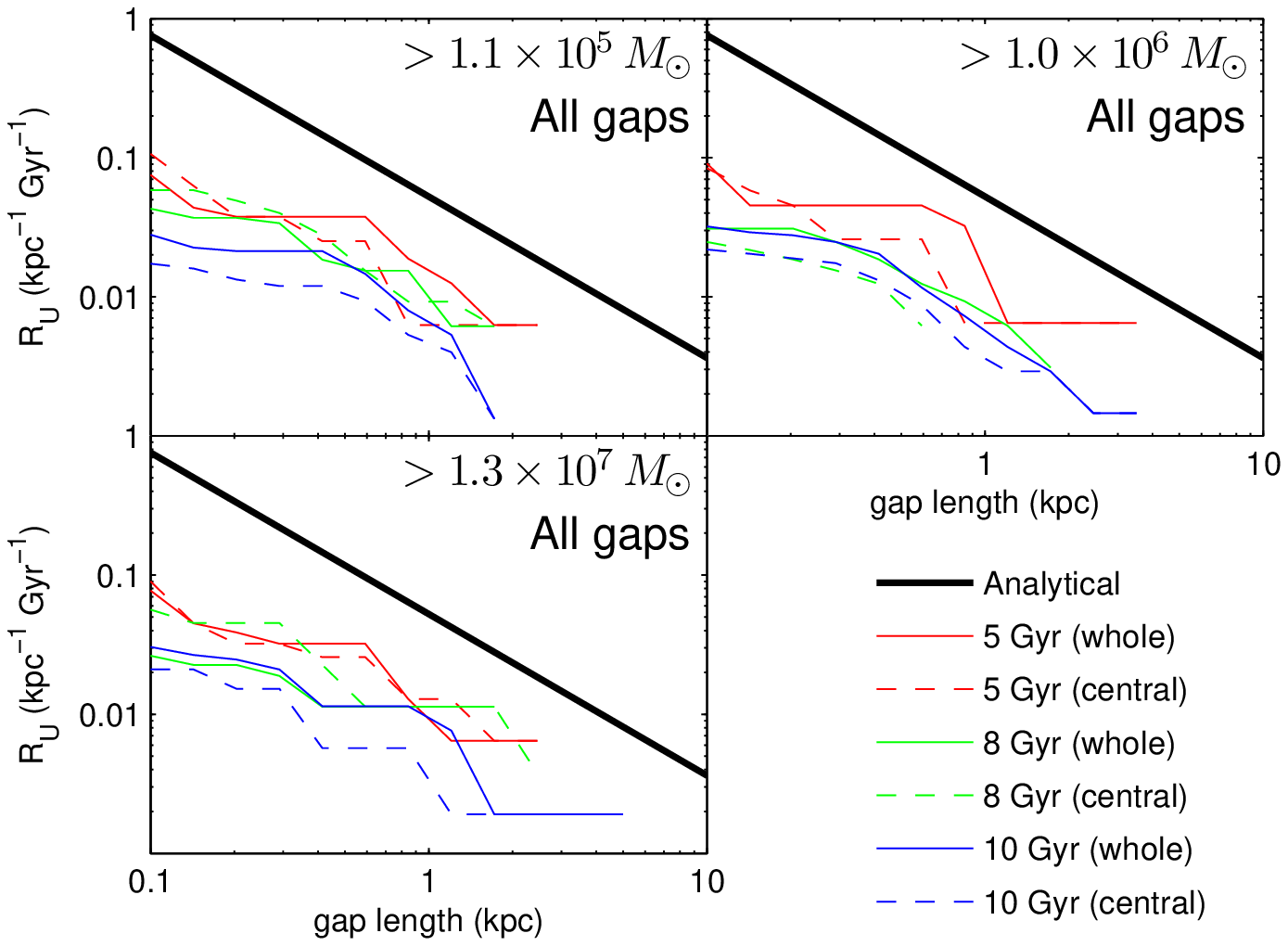}
	\includegraphics[width=3.5in]{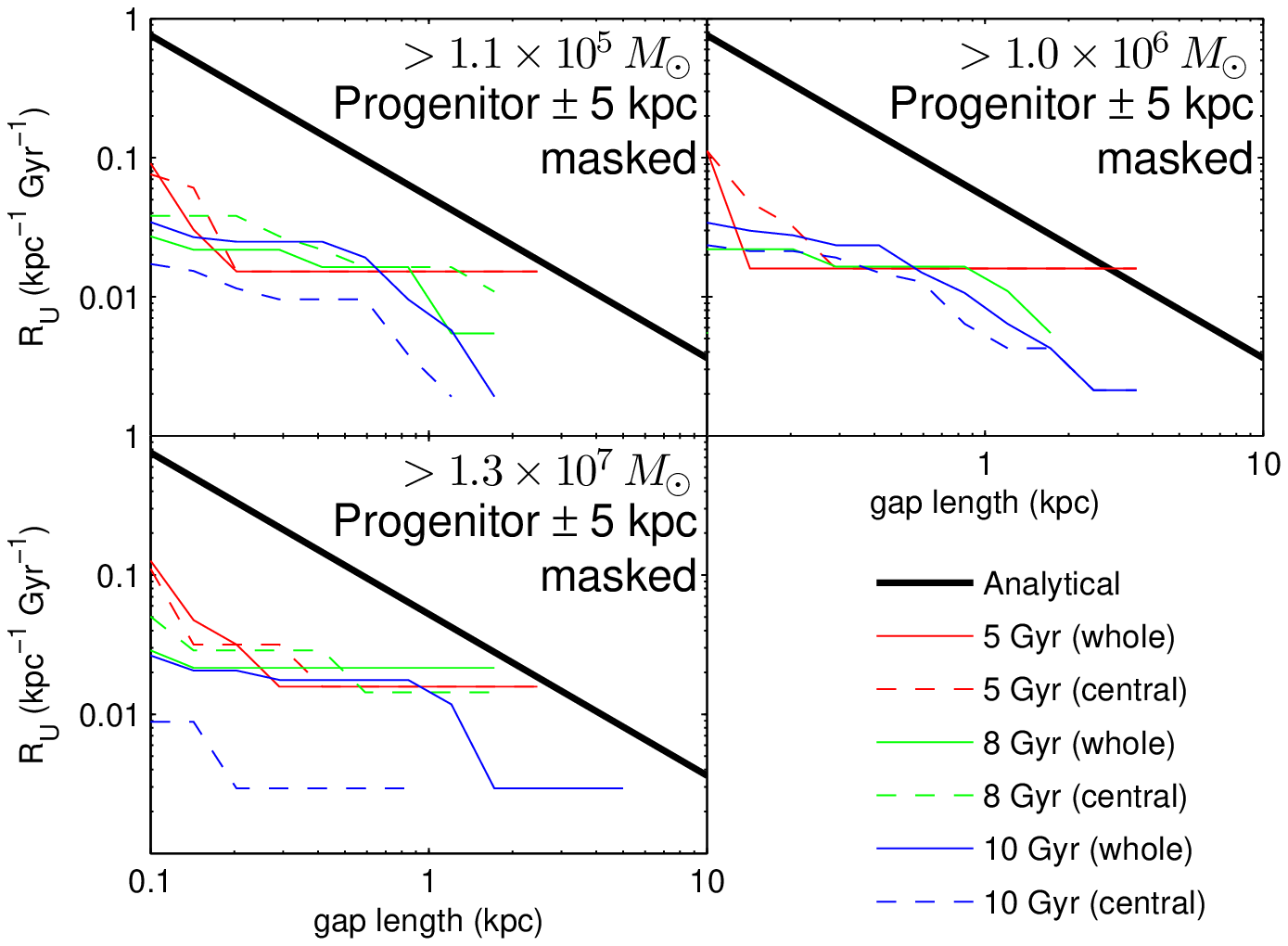}
	\caption{Analytical \citep{carlberg12} and simulated gap spectra. The gaps are collected from single simulations which include the full ranges of
	subhalos. The lower limit of the mass range is labeled in each panel, and the upper limit is $10^8\Msun$ in all cases.
	In the top three panels, all gaps identified (including EOs) \newrevision{are} included in gap spectrum. In the bottom three panels, the gaps
	located within $5\,\mathrm{kpc}$	away from the progenitor \newrevision{are} eliminated, and the stream lengths and ages \newrevision{are} adjusted.
	\newrevision{Compared to \fig{RU_summed}, the simulated gap spectra here no longer follow the analytical prediction because
	of overlapping between the gaps.}}
	\label{fig:RU_all}
\end{figure}

We now consider the validity of \eqn{gaprate_gapsize_formula} for stream gaps in the presence of all subhalos
in each cumulative mass range in \tab{subhalo_massrange}. \fig{RU_all} compares the measured gap spectra of both
the whole and central streams for three mass ranges, with and without the gaps within $5\,\mathrm{kpc}$ away
from the progenitor. Clearly in all cases, the ideal gap spectrum over estimates the measured spectrum by nearly an order
of magnitude due to gap overlapping.

Similar to the simulations with independent sets of subhalos (\fig{RU_summed}), the gap spectra produced by full sets
of subhalos do not have strong dependence on stream age and integrating width. The only exception is the youngest stream
shown at 5 Gyr which consistently has higher $R_\cup$ than the older streams. However, when the gaps near the progenitor
are eliminated, the numbers of gaps at $5\,\mathrm{Gyr}$ in all cases decrease significantly, where the gap spectra are
dominated by a single gap at $2-3\,\mathrm{kpc}$, and a number of extremely short gaps. This is likely because the stream
is still young, and the effective length of the stream (after masking $10\,\mathrm{kpc}$ centered at the progenitor) is only
$\sim15\,\mathrm{kpc}$. This eliminates a significant part of the stream, making its stream statistics unreliable.

The weak dependence of the gap spectrum on the integrating width for linear density is also worth noting. \fig{stream_maps_lcdm}
shows that gaps in general have much more complicated morphologies than straight edges across the width of the stream.
The explanation for these morphologies requires detailed understanding of how subhalo perturbations manifest in a 
self-consistent stream, which is beyond the scope of this study. While \citet{carlberg13} studied the dynamics of subhalo
perturbations for an idealized stream, we defer the self-consistent case to a future study.

Perhaps the most surprising result is that the gap spectra do not show obvious dependence on subhalo masses. The spectra
are difficult to distinguish between the mass range of subhalos which causes the gaps. This is in
disagreement with \citet{carlberg12} which derived a relation between the length of a gap and the mass of the
subhalo that caused it such that
\begin{equation}
	l(M) = 8.3 \left( \frac{r}{30\,\mathrm{kpc}} \right)^{0.37} \left( \frac{M}{10^8\Msun} \right)^{0.41}\,\mathrm{kpc}.
	\label{eqn:length_mass}
\end{equation}
From this formula, it is reasonable to expect the inclusion of lower-mass subhalos to show more gaps at the shorter end. 
In their ideal simulations, however, \citet{carlberg12} did not account for the time evolution of each gap. 
An example can be seen in \figs{density_gaps_whole_massrange13}{density_gaps_central_massrange13}. The gap located at
about $-7\,\mathrm{kpc}$ at $4\,\mathrm{Gyr}$ evolves into a much longer gap centered at about $-9\,\mathrm{kpc}$ at
$10\,\mathrm{Gyr}$. Evidently \eqn{length_mass} requires revision for self-consistent streams before it can be used
to understand the relation between gap spectra and subhalo masses.

\subsection{Observational Considerations}
\label{sec:case_study}

\newrevision{We now consider the issues when interpreting gap spectra from observations. A gap spectrum for GD-1 has been
observed by \citet{CG13}, but we emphasize that the gap spectra from our simulated streams in this study should not be directly
compared to the one in \citet{CG13} because our models for both the star cluster
and the galaxy halo are chosen in favor of a simple interpretation, and may be missing some complications discussed
in Section \ref{sec:other_effects}.}

\newrevision{We first project each stream onto sky coordinates. For
simplicity, we put the hypothetical observer at the center of the galaxy, and then project each
particle onto the azimuthal and altitudinal plane in galactocentric coordinates. Since the our stream progenitor is
orbiting along the $xy$-plane in a spherical potential, the smooth stream appears as a straight
line along the azimuthal direction, and each \lcdm\ stream appears only a few degrees off the azimuthal
plane due to subhalo perturbations.}

\newrevision{The density along the stream is simply the number of particles in bins of $0.1\degrees$ in the 
azimuthal direction. The match filter approach to detect gaps
remain the same as the analysis above, but the 12 filter scales (\fig{filter_scales}) are now logarithmically spaced
in angular units from $0.34\degrees$ to $14\degrees$, and the noise levels are obtained from the regions at
$>10\degrees$, rather than $>5\,\mathrm{kpc}$, away from the progenitor. The choice of bin size and filter scales 
are on the same orders of magnitude as \citet{CG13}, but putting the hypothetical observer
at the Galactic center may affect angular sizes by factors of $\sim2$.}

\newrevision{To ensure that the behaviors of the simulated streams are typical, we simulate each \lcdm\ stream 10 times
with the same initial conditions for the star cluster, but different realizations of the same subhalo distributions.
At the end we take the median numbers of gaps of the 10 streams to avoid outliers.}

\subsubsection{Orbital Phase}

\begin{figure}
	\centering
	\includegraphics[width=3.5in]{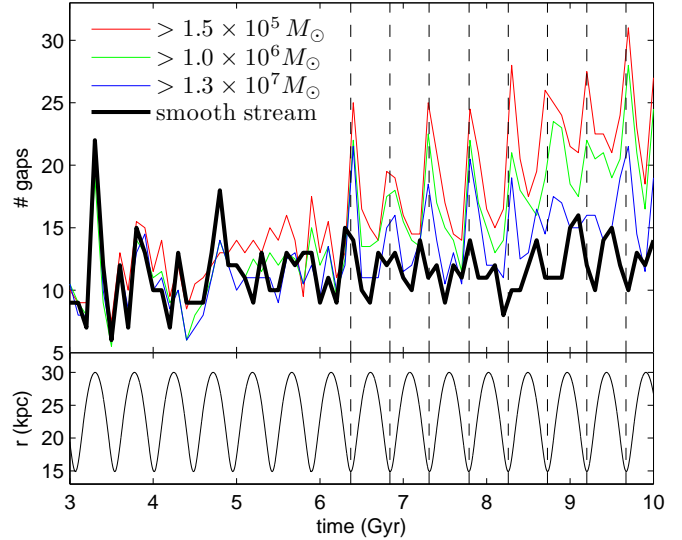}
	\caption{\newrevision{Top: time evolution of the number of gaps longer than $0.34\degrees$.
	Each colored line represents the median of
	10 realizations of an identical subhalo distribution (with $10^8\Msun$ as upper mass limit),
	and the black thick line represents the smooth stream. Bottom: the
	vertical dashed lines are visual guides which show that the variation of the number of gaps are correlated with 
	the orbital phase of the stream progenitor. These plots show that the
	cumulative number of gaps in a stream on average is increasing in time, but instantaneously the number of gaps
	observable can have an even stronger dependence on the orbital phase of the stream than
	on the stream's dynamical age or subhalo masses.}}
	\label{fig:numgaps_time}
\end{figure}

\newrevision{One surprising result from Section \ref{sec:lcdm_stream_with_all_subhalos} is that the gap spectrum has little to no dependence
on age and subhalo masses. To investigate what this means when interpreting observations, the top panel of \fig{numgaps_time} shows
the cumulative numbers of gaps longer than $0.34\degrees$ (i.e., all gaps detected in the entire stream) as
functions of time. At $t>5\,\mathrm{Gyr}$,  
the numbers of gaps due to subhalos vary according to the orbital phase of the stream progenitor. The ``bursts'' in numbers of gaps 
in the \lcdm\ streams occur when the streams are stretched as they passes through the pericenters of their orbits.
At $t<5\,\mathrm{Gyr}$, on the other hand, this correlation does not exist for two reasons. First, our detection
method (Section \ref{sec:gapfinding}) uses the parts of the stream that are $>5\,\mathrm{kpc}$ (before sky projection)
or $>10\degrees$ (after sky projection) away from the progenitor in order to estimate noise. At $t<5\,\mathrm{Gyr}$,
the length of the stream varies between a few to $20\,\mathrm{kpc}$, which may not be long enough to estimate noise.
Second, in only $5\,\mathrm{Gyr}$ the stream does not yet have enough time and to interact with subhalos.}

%At $t\lesssim5\,\mathrm{Gyr}$, subhalos have not made a significant contributions to the gaps, and 
%the most of the detected gaps are EOs. As subhalos begin to dominate the number of gaps at $t>5\,\mathrm{Gyr}$, we find 
%that these gaps vary according to the orbital phase of the stream progenitor. The ``bursts'' in numbers of gaps 
%in the \lcdm\ streams occur when the streams are stretched as they passes through the pericenters of their orbits.
%The gaps due to EOs, on the other hand, are insensitive to orbital phases.}

\newrevision{Both the total number of gaps and the dynamical age of the stream are difficult to measure,
as the some parts of a stream may not be observable.
We define a more useful quantity $S_\cup$ which is the cumulative number of gaps longer than a given gap length per
unit stream length. In other words, leaving the age of the stream as an unknown, $S_\cup$ differs 
from $R_\cup$ in Section \ref{sec:gap_size_distribution} by a normalization by age, and that $S_\cup$ is after sky projection.
In the next section we show that gaps due to EOs and subhalos have very different $S_\cup(l)$ distributions which 
are directly observable.
}

%\newrevision{Although the middle panel of \fig{numgaps_time} shows that the total gap formation rates of EOs and subhalos
%are similar, in the next section we show that these two types of gaps have very different $S_\cup(l)$ distributions.}

\subsubsection{Signal-to-noise Ratios}

\newrevision{Another important distinction between simulations and observations is the signal-to-noise ratio (S/N).
At 5 and $8\,\mathrm{Gyr}$ our smooth stream is represented by about 60,000 and 80,000 particles in total, respectively (\fig{massloss}).
\citet{koposov10} estimated that the $60\degrees$ visible segment of GD-1 consists of 3000 stars. At an average
distance of $\sim10\,\mathrm{kpc}$, the visible segment is $\sim10\,\mathrm{kpc}$ long. In our simulations,
after 5 and $8\,\mathrm{Gyr}$, the average stream lengths are about 20 and $40\,\mathrm{kpc}$, respectively (\fig{massloss}).
This means that our simulated stream should be represented about eight times fewer particles in order to be comparable to observations.
With the progenitor masked, we reduce the number of particles in the stream by randomly sampling the stream using
two, four, and eight times fewer particles than the original stream. The particle reduction applies to both the stream of interest
and the smooth stream which is the source for estimating noise. This allows us to investigate the importance of 
high S/N.}

\newrevisionII{Note that in our simulation each particle is equivalent to about $0.043\Msun$, which is less massive than
the typical stars that are detected in observations. Our simulations are not meant to be physical models of the real stream.
In this section, we are only concerned about matching the numbers of particles in the simulations to the numbers of stars
in the observation. As the stars escape from the progenitor, the stream's self-gravity becomes negligible \citep{johnston98},
and the particles' masses are no longer important.}

\begin{figure}
	\centering
	\includegraphics[width=3.5in]{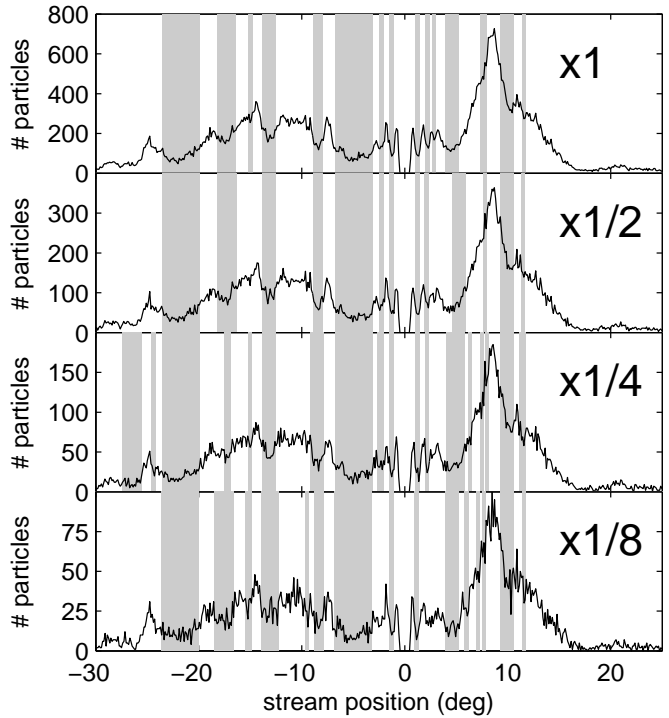}
	\caption{\newrevision{Density profiles and gaps detected at 99\% confidence for the \lcdm\ stream with subhalos of masses
	$1.5\sci{5}<M/\Msun<10^8$ at 8 Gyr, projected onto the sky. Each panel represents a stream whose number of
	particles have been reduced by the factor indicated and binned at $0.1\degrees$ throughout. Without reduction
	(top panel), the stream contains about 80,000 particles. The stream retains most of its gaps even after particle reduction 
	by a factor of eight.}}
	
%	The stream retains most of its gaps after a reduction by a factor of 4, but loses a significant
%	number after a reduction by a factor of 10.}}
	\label{fig:reduction_gaps}
\end{figure}

\newrevision{\fig{reduction_gaps} shows the density profiles of the \lcdm\ stream with subhalo masses $1.5\sci{5}<M/\Msun<10^8$
at $8\,\mathrm{Gyr}$ projected onto the sky.
The panels show the gaps detected in the same stream after three levels of particle reduction.
Even after reduction by a factor of eight, the stream appears to have retained most of its gaps despite a lower
S/N.}

%\newrevision{\fig{reduction_gaps} shows the density profiles of the \lcdm\ stream with subhalo masses $1.5\sci{5}<M/\Msun<10^8$
%at $8\,\mathrm{Gyr}$ projected onto the sky.
%The panels show the gaps detected in the same stream after three levels of particle reduction.
%After reduction by factors of 2 to 4, the stream retains most of the gaps. After reduction by
%a factor of 10 ($\sim8,000$ particles for the whole stream), however, 
%the signal-to-noise ratio decreases to the point where a lot of gaps are no longer detected.
%Because of the small number of gaps at low signal-to-noise ratios, we repeat the simulation nine more times
%with identical initial condition for the star cluster, but different random seeds for the same subhalo 
%distribution.}

%\newrevision{Neither the dynamical age nor the entire angular extent of GD-1 are well known. \citet{CG13} reports
%their gap spectrum as the number of gaps observed in the $60\degrees$ segment of the stream. We use a slightly
%modified gap spectrum $S_\cup$ which is the cumulative number of gaps longer than gap length $l$ per unit
%stream length. In other words, $S_\cup$ and $R_\cup$ only differs by stream age, which is left as an unknown.
%In addition, we also reduce the number of gaps in \citet{CG13} by a factor of 1.44, as discussed in their
%\S 3.3.}

\begin{figure*}
	\centering
	\includegraphics[width=6.5in]{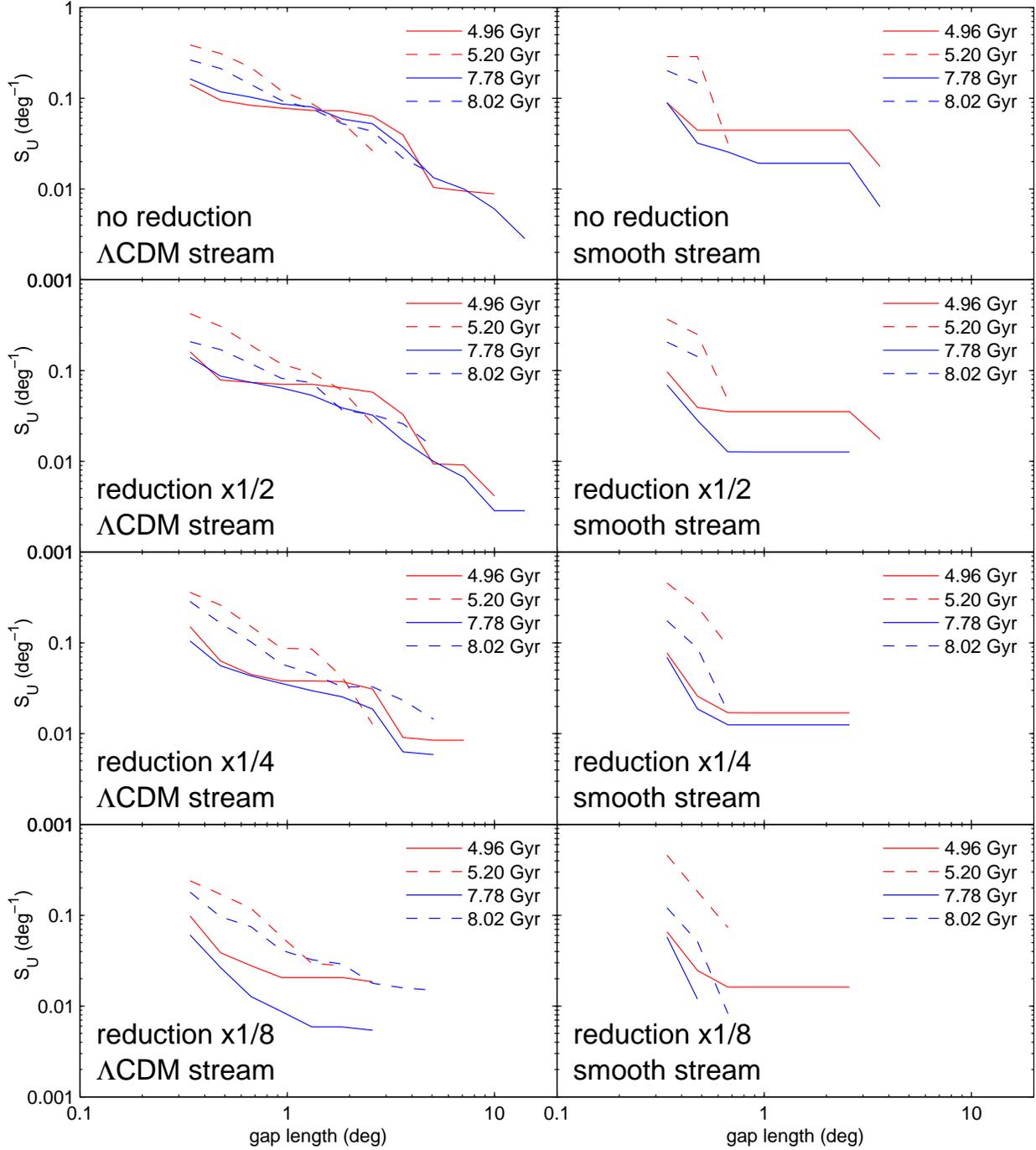}
	\caption{\newrevision{Gap spectra from simulated streams after sky projection. 
	Each line in the panels on the left are the median of 10 spectra of the same \lcdm\ stream but with ten identical subhalos
	distribution of masses $1.5\sci{5}<M/\Msun<10^8$. The panels on the right are the spectra of the smooth stream. The 
	spectra of the \lcdm\ stream show presence of gaps longer than $1\degrees$, whereas the spectra of the 
	smooth stream quickly drops off. Colored solid (dashed) lines are spectra obtained when the stream is undergoing
	pericentric (apocentric) passage. Each row represents the spectra after the numbers of particles in the streams
	have been reduced by the factor labeled.  The bottom panels indicate that a high S/N detection of the stream
	is important for understanding the origins of gaps, otherwise the gap formation rates can be systematically 
	biased depending on the stream's orbital phase.}}
%	\caption{Comparing the gap spectra from simulations and observations of GD-1 taken from \citet{CG13}. Black
%	solid line is the observation. Colored solid lines are gap spectra in the \lcdm\ stream with subhalos of masses 
%	$6.5\sci{4}<M/\Msun<10^8$ at various ages, and colored dashed lines are the gap spectra in the smooth stream with
%	no subhalos. The \lcdm\ stream at all ages shows a tail for gaps at a few degrees long, whereas the smooth stream
%	does not have gaps longer than $1\degrees$.}
	\label{fig:GD1_reduction}
\end{figure*}

\newrevision{In \fig{GD1_reduction}, each line shows the median of ten gap spectra from the same stream but with 10
realizations of the same subhalo distribution. In each panel, the solid (dashed) lines represent the times
when the progenitor is at the pericenter (apocenter) of its orbit. When the stream is compressed and stretched
as it oscillates radially (see \fig{massloss}), its length can differ by up to 
a factor of two. Careful inspection of \figs{numgaps_time}{GD1_reduction} shows that during pericentric passages,
the numbers of gaps are at maximum, but $S_\cup$ is at minimum because the stream length is also at maximum.
For a \lcdm\ stream at high S/N (upper left panel in \fig{GD1_reduction}),
the gap spectra are not sensitive to this oscillation, except with an excess of shorter gaps and fewer longer
gaps, which are expected as the stream, including its longitudinal structure, is compressed during apocentric passage.
At low S/N
(lower left panel in \fig{GD1_reduction}), however, the gap spectrum during apocentric passage is consistently
higher than that during pericentric passage. This is because the length of the stream is insensitive to 
the S/N, but the number of gaps is not. Therefore, high S/N data for the stream
is important when studying stream gaps. Otherwise, the spectrum may be over- or underestimated depending on 
the orbital phase.}

\newrevision{The right panels in \fig{GD1_reduction} show the gap spectra of the smooth stream. They are also somewhat sensitive
to S/N, but the most obvious difference from the spectra of \lcdm\ stream is the shape of
the spectra. This is especially obvious during pericentric passage where the gap spectra
rapidly drop off to zero for gaps longer than $\sim1\degrees$. Even during apocentric passage, the gap spectra remain flat
at gap lengths $\gtrsim1\degrees$. If the gaps originated from subhalo perturbations, then the gap spectrum
should be steep and extend well beyond $1\degrees$.}

\subsubsection{The Case of GD-1}

\newrevision{
An interesting confusion for GD-1 in particular is that GD-1's progenitor has not been identified. If GD-1's
progenitor has evaded observation, and the observation corresponds to a segment of the stream which is
close enough to the progenitor such that EOs can be observed, then GD-1 may be a poor choice
as a probe for missing satellites. However, this is unlikely because EOs are observable only
in a small segment of the stream, and subhalo gaps are observable everywhere along the stream.
}

\newrevision{
Another possibility is that GD-1's progenitor may have been completely disrupted. In this case, the tidal radius
of the progenitor approaches zero. Since the spacing between EOs are proportional to the
tidal radius of the progenitor, this means that the EOs should also fade away as the progenitor
is disrupted \citep{kupper10}. Therefore, despite its lack of progenitor GD-1 should be a viable probe for missing satellites.
}

\newrevision{
The gap spectrum of GD-1 has been measured by \citet{CG13}. The spectrum show presence of gaps at all lengths
between $0.2\degrees$ and $10\degrees$, which is sufficient to rule out a smooth and spherical potential.
To facilitate a conclusive analysis on the origin of the gaps in GD-1, we need to consider a much more realistic
model which include the effects discussed in Section \ref{sec:other_effects}.
}

\subsection{Subhalo Mass Limits}

We consider the effects of subhalos as massive as $10^8\Msun$, since the effects by more massive subhalos are not
relevant to us. Gaps caused by subhalos at these masses produce 
long gaps with high-density contrasts. For example, an obvious gap located at $-7\,\mathrm{kpc}$ at 4 Gyr shown in
\figs{density_gaps_whole_massrange13}{density_gaps_central_massrange13} are caused by a $4.5\sci{7}\Msun$
subhalo. In fact,
the perturbation by $M\gtrsim10^8\Msun$ ($h\gtrsim 1\,\mathrm{kpc}$) subhalos can be so catastrophic that the stream
is warped and divided into segments. As a result, a stream which originated from one progenitor can be observed
as a few separate streams. Observations of Pal-5's and GD-1's gaps, on the other hand, show small-scale density fluctuations in 
a long, narrow stream, so these two streams are not sensitive to subhalos above $10^8\Msun$. By coincidence, this
upper limit approximately coincides with the upper limit beyond which the models of warm dark matter can be no longer be distinguished from
CDM. Therefore, $10^8\Msun$ is a reasonable upper limit where our simulations 
can be useful.

In the low-mass end, we only consider the effects of subhalos down to $\sim 6\sci{4}\Msun$. From a separate simulation of the 
same stream but with only the subhalos with masses $6.5\sci{4}<M/\Msun<7.5\sci{4}$, the density profile is indistinguishable
from the smooth stream, and the gap statistics are identical. Furthermore, Section \ref{sec:lcdm_stream_with_all_subhalos} shows
that the gap spectra have very little dependence on mass. Changing the mass lower limit from $7.5\sci{4}\Msun$ to 
$6.5\sci{4}\Msun$ produced indistinguishable gap spectra. 
This means that subhalos less massive than $\sim10^5\Msun$, even though they are much more abundant than those of
higher masses (\eqn{subhalo_massfunction}), have minimal effects on our stream.

\newrevision{The Milky Way has about 160 known GCs \citep{harris}, and a few hundred DGs brighter than
$L\gtrsim10^3 L_\odot$ after bias corrections (see \citet{bullock10} for a review). It is interesting to ask whether these
known satellite systems, rather than the truly ``missing'' satellites, can contribute
to the observed stream gaps. Typical GCs have masses $\sim10^5\Msun$, which correspond to the low
end of our mass spectrum of subhalos. In the same mass range, though, there are orders of magnitudes more subhalos
(e.g., $\sim10^5$ subhalos at $10^5<M/\Msun<10^6$) than GCs, so GCs are unlikely to contribute significantly to observed
stream gaps. On the other hand, DGs are commonly found at $\gtrsim10^7\Msun$ \citep{strigari08} which is the high
end of our mass spectrum of subhalos. At that mass range ($\sim2000$ subhalos at $M>10^7\Msun$), the number of known
DGs are only 1 order of magnitude below the number of subhalos, so
DGs may contribute to some observed gaps. However, a common limitation in understanding the contributions from both
GCs and DGs is their orbits, especially when the kinematics of these satellites are not well constrained.
As done in our simulations (Section \ref{sec:subhalo_mass_ranges}), subhalos that do not
approach the stream's orbit will interact minimally with the stream. \tab{subhalo_massrange} shows that in our
realizations of subhalos, only $\sim15\%$ of them would approach to within $2\,\mathrm{kpc}$ of GD-1's orbit. 
This means that most known satellites may never interact with a GD-1-like stream, and that stream gaps, if they
were indeed due to satellites and were not EOs, are more likely due to satellites that have never been observed.
}

\subsection{Other Effects}
\label{sec:other_effects}

In order to keep our results simple, the galaxy is modeled as stationary, spherical NFW potential, 
the subhalos as static, test masses, and the satellite as a collisionless King model. These models ignore a
number of known dynamical complications.

{\it Two-body Relaxation.} The star cluster is modeled as a collisionless system with relaxation timescale
of $\sim 110\,\mathrm{Gyr}$.
Globular clusters typically have relaxation timescales of $\lesssim10\,\mathrm{Gyr}$ \citep[2010 Edition]{harris},
so mass loss should originate from dynamical evaporation, in addition to tidal disruption. As a result,
the star cluster should be disrupted even faster than we measured in \fig{massloss}. This may have an important
effect on the formation of gaps, which depends on the details of the dynamics of a stream \citep{carlberg13}.
The relation between gaps and mass loss mechanism will be investigated in a future study.

{\it Dynamical Friction (DF).} Both the star cluster and subhalos should suffer from DF as they orbit
around the \newrevision{dark matter halo}. Comparing the magnitudes of the accelerations due to DF and due to the orbit,
$a_{DF}/a_{orbit}\sim10^{-8}\ln\Lambda$ for both the star cluster at 22 kpc and a $10^6\Msun$ subhalo at 100 kpc,
where $\ln\Lambda\equiv \ln(b_{max}/b_{min})\approx 10$ is the log of the ratio of the maximum and minimum impact distances
\citep{bt08}. Therefore, DF is negligible throughout our model.

{\it Disk Shocking:} \citet{dehnen04} found that the evolution of Pal-5 is driven by the tidal shocks when
crossing the Galactic disk, which is not modeled in our simulations. The orbit of Pal-5 in \citet{dehnen04}
has peri- and apogalacticon at 5.5 kpc and 19 kpc, respectively, whereas our smooth stream has peri- and
apogalacticon at 15 kpc and 30 kpc, respectively. Being farther away from the the Galactic center, if our simulations
contained a disk, its effect should be less severe for our simulated stream than Pal-5. 
Moreover, \citet{dehnen04} concluded that disk shocking is
not responsible for the observed structure in Pal-5, while \citet{kupper10} concluded that EOs persist even
under the influence of disk shocks, so the absence of a disk should not significantly change our conclusion.
The gap formation rate with and without subhalos in the presence of a disk is beyond the scope of this study.

{\it Halo Shape and Collapse History:} \citet{SV08} found that 
\newrevision{the shape of the halo potential can have a larger effect than subhalos have on the over all
structure of a stream. However, their simulations focused on streams which originated from DGs at
$\sim10^9\Msun$, as well as subhalos at $\gtrsim10^7\Msun$ which is the high end of our mass spectrum of
subhalos. The small gaps from a stream originating from a GC in a non-spherical halo has yet to be studied.}
In fact, since the initial collapse of the entire halo, the potential cannot be stationary throughout a Hubble time,
which is the timescale of our simulations. In the future,
\newrevision{we aim to repeat a similar study using potentials which resulted directly from high resolution simulations
such as \citet{madau08} and \citet{aquarius}. The self-consistent halo and subhalo potentials from those simulations can eliminate the idealized
models in Section \ref{sec:models}.}

\section{Conclusions}

For the first time, we used N-body simulations to model the disruption of
a collisionless star cluster which formed a narrow stream similar to Pal-5 and GD-1, and
we investigated the phenomenology of gaps that originated from the perturbations by subhalos predicted in the \lcdm\ cosmological
model. Analytical predictions of stream gap statistics in previous studies
were all based on massless particles distributed to mock realistic streams, but the dynamics of gaps have
never been studied in self-consistent models. With a stream from a self-consistent model,
we characterized the gap length distribution which can be used as a tool to understand the origin
of stream gaps seen in observations.

%The properties of the subhalos in our simulations were taken from Aquarius simulation \citep{aquarius}.
%We divided the subhalos into thirteen mass ranges
%from $6.5\sci{4}\Msun$ to $10^8\Msun$, and we ran fourteen simulations of the same stream -- one without any
%subhalos (the smooth stream), and one for each cumulative mass range (the \lcdm\ streams). In each stream we looked for gaps
%using a matched filter approach previously used by \citet{pal5_gaps,CG13}.

The properties of the subhalos in our simulations were approximations to those in the Aquarius simulation \citep{aquarius}.
We ran 14 simulations of the same stream -- 1 without any
subhalos (the smooth stream), and 1 for each cumulative mass range in \tab{subhalo_massrange} (the \lcdm\ streams).
In each stream we looked for gaps using a matched filter approach previously used by \citet{pal5_gaps} and \citet{CG13}.
We found that, in addition to subhalo perturbations, the overdensities of particles due to their epicyclic motions
as the progenitor loses mass \citep{kupper08} can also produce gaps. Therefore, even without subhalos,
``gaps'' can appear within $\sim5\,\mathrm{kpc}$ away from the progenitor.
\newrevision{For the first time, our match filter approach accounted for these EOs together
with the gaps due to subhalo perturbations.}

%Our match filter approach accounted for these epicyclic overdensities (EOs), and we compared the gap spectra, which are
%the gap formation rates as functions of gap length, with and without the presence of subhalos.

\citet{yoon11} first noted that subhalos gaps were typically diagonal and not perpendicular to the stream due to 
the range in angular momenta across the width of the stream. 
We investigated whether this could be a hindrance to gap detection. By measuring the distribution of angular momenta in our simulated stream, 
we estimated that the two ends of a gap across the width of a stream were sheared by no more than a $1\,\mathrm{kpc}$ per Gyr.
Rather, subhalo gaps show complicated morphologies which were already imprinted into the stream as soon as the
\newrevision{gaps first} occurred. 
In addition to integrating the entire thickness of the stream, we also considered the case where the linear density
are integrated using only the central 0.08 kpc of the stream in order to minimize the impact of gaps morphologies.
We found that the resulting gap rate spectra the two cases were similar. Therefore, gap morphology does not affect our conclusion.

We tested the validity of the idealized gap spectrum $R_\cup$, or the cumulative number of gaps per unit stream length
per unit stream age as a function of gap length $l$, derived by \citet{carlberg12}. We found that overlapping gaps in the 
stream can significantly reduce $R_\cup$, \newrevision{and that the dependences of $R_\cup$ on subhalo masses and stream
age are smaller than its dependence on the stream's orbital phase. Therefore, the stream's orbital phase must be known
when interpreting gap formation rates in observations.}

\newrevision{We considered how to interpret gap spectra from observations by projecting the stream onto the sky, and
for each \lcdm\ stream we also simulated them using ten realizations of the same subhalo distributions. One observational
concern is the S/N of the stream's detection. We down-sampled our simulated streams with less particles in order
to match the S/N which is similar to the GD-1 detection \citep{CG13}. Our result indicated that 
at GD-1's S/N, the gap spectrum can be biased by the orbital phase of the stream. In addition, we compared gap spectra
produced purely by EOs and by EOs and subhalos together in a \lcdm\ halo. We showed that the gap spectra of the former
are limited in gap lengths, and that the latter have a much larger variety of gap lengths. This can be a powerful method
to identify the origin of gaps in streams. Therefore, high S/N data such as
those from {\it Gaia} will be very useful for using stream gaps to constraint the abundance of subhalos.
}

The dynamics of stream gaps depend on the details of the dynamics of the stream itself. We adapted a few tools such
as match filter and scaling relations which were derived from idealized simulations. In a future study, we aim to 
use self-consistent streams to repeat experiments akin to \citet{yoon11}, \citet{carlberg12}, and \citet{carlberg13},
where the details of individual gaps can be studied in controlled experiments, in order to revise the aforementioned tools
that is applicable quantitatively to realistic streams.

\newrevision{We thank the anonymous referee for valuable comments, which inspired us to expand
the content of this paper. Computations were performed on the GPC supercomputer at the SciNet HPC Consortium.
SciNet is funded by: the Canada Foundation for Innovation under the auspices of Compute Canada,
the Government of Ontario; Ontario Research Fund - Research Excellence, and the University of Toronto.}

%\begin{figure}
%	\centering
%	\includegraphics[width=3.5in]{figs/gaprate_gapsize/gaprate_gapsize_GD1.eps}
%	\includegraphics[width=3.5in]{figs/gaprate_gapsize/gaprate_gapsize_GD1_mass1.1e5_1e8.eps}
%	\includegraphics[width=3.5in]{figs/gaprate_gapsize/gaprate_gapsize_GD1_mass6.5e4_1e8.eps}
%\end{figure}

%\clearpage

%% Tables may also be prepared as separate files. See the accompanying
%% sample file table.tex for an example of an external table file.
%% To include an external file in your main document, use the \input
%% command. Uncomment the line below to include table.tex in this
%% sample file. (Note that you will need to comment out the \documentclass,
%% \begin{document}, and \end{document} commands from table.tex if you want
%% to include it in this document.)

%% \input{table}

%% The following command ends your manuscript. LaTeX will ignore any text
%% that appears after it.


\begin{thebibliography}{}

\bibitem[Bardeen et al. (1986)]{bardeen86} Bardeen, J. M. Bond, J. R., Kaiser, N., et al. 1986, \apj, 304, 15

\bibitem[Barkana et al. (2001)]{barkana01} Barkana, R., Haiman, Z., \& Ostriker, J. P. 2001, \apj, 558, 482

\bibitem[Benson et al. (2013)]{benson13} Benson, A. J., Farahi, A., Cole, S. 2013, \mnras, 428, 1774

\bibitem[Binney \& Tremaine (2008)]{bt08} Binney, J. \& Tremaine, S. 2008, Galactic Dynamics (2nd ed.; Princeton, NJ: Princeton Univ. Press)

\bibitem[Blumenthal et al. (1984)]{blumenthal84} Blumenthal, G. R., Faber, S. M., Primack, J. R., et al. 1984, \nat, 311, 517

\bibitem[Bode et al. (2001)]{bode01} Bode, P., Ostriker, J. P., \& Turok, N. 2001, \apj, 556, 93

\bibitem[Bullock (2010)]{bullock10} Bullock, J. S. 2010, arXiv:1009.4505

\bibitem[Carlberg (2009)]{carlberg09} Carlberg, R. G. 2009, \apjl, 705, L223

\bibitem[Carlberg (2012)]{carlberg12} Carlberg, R. G. 2012, \apj, 748, 20

\bibitem[Carlberg (2013)]{carlberg13} Carlberg, R. G. 2013, \apj, 775, 95

\bibitem[Carlberg \& Grillmair (2013)]{CG13} Carlberg, R. G., \& Grillmair C. J. 2013, \apj, 768, 171

\bibitem[Carlberg et al.(2012)]{pal5_gaps} Carlberg, R. G., Grillmair C. J., \& Hetherington N. 2012, \apj, 760, 75
    
\bibitem[Comparetta \& Quillen (2011)]{qc11} Comparetta, J., \& Quillen, A. C. 2011, \mnras, 414, 810

\bibitem[Davis et al. (1985)]{davis85} Davis, M., Efstathiou, G., Frenk, C. S., et al. 1985, \apj, 292, 371

\bibitem[Dehnen et al. (2004)]{dehnen04} Dehnen, W., Odenkirchen, M., Grebel, E. K., et al. 2004, \aj, 127, 2753

\bibitem[Eyre \& Binney (2011)]{EB11} Eyre, A., \& Binney, J. 2011, \mnras, 413, 1852

\bibitem[Gao et al. (2011)]{gao11} Gao, L., Frenk, C. S., Boylan-Kolchin, M., et al. 2011, \mnras, 410, 2309

\bibitem[Grillmair (2010)]{G10} Grillmair, C. J. 2010, in Galaxies and Their Masks, ed. D. Block, K. C. Freeman, \& I. Puerari (New York: Springer), 247

\bibitem[Grillmair \& Dionatos (2006)]{gd1} Grillmair C. J., \& Dionatos O. 2006, \apjl, 643, L17

\bibitem[Harris (1996) (2010 Edition)]{harris} Harris, W. E. 1996, \aj, 112, 1487

\bibitem[Helmi et al (2003)]{helmi03} Helmi A., White S. D. M., \& Springel, V. 2003, \mnras, 339, 834

\bibitem[Hernquist (1990)]{hernquist90} Hernquist, L. 1990, \apj, 356, 359

\bibitem[Ibata et al. (2002)]{ibata02} Ibata, R. A., Lewis, G. F., Irwin, M. J., et al. 2002, \mnras, 332, 915

\bibitem[Johnston (1998)]{johnston98} Johnston, K. V. 1998, \apj, 495, 297

\bibitem[Just et al. (2009)]{just09} Just, A., Berczik, P., Petrov, M. I., et al. 2009, \mnras, 392, 969

\bibitem[Kamionkowski \&  Liddle (2000)]{kamionkowski00} Kamionkowski, M., \& Liddle, A. 2000, PhRvL, 84, 4525

\bibitem[Klypin et al. (1999)]{klypin99} Klypin, A., Kravtsov, A. V., Valenzuela, O., et al. 1999, \apj, 522, 82

\bibitem[Koposov et al. (2010)]{koposov10} Koposov, S. E., Rix H. W., \& Hogg D. W. 2010, \apj, 712, 260

\bibitem[Koposov et al. (2009)]{koposov09} Koposov, S. E., Yoo, J., Rix, H.-W., et al. 2009, \apj, 696, 2179

\bibitem[K\"upper et al. (2008)]{kupper08} K\"upper, A. H. W., MacLeod, A., \& Heggie D. C. 2008, \mnras, 387, 1248

\bibitem[K\"upper et al. (2010)]{kupper10} K\"upper, A.H.W., Kroupa, P., Baumgardt, H., et al. 2010, \mnras, 401, 105

\bibitem[Law et al. (2005)]{law05} Law D. R., Johnston K. V., \& Majewski S. R. 2005, \apj, 619, 807

\bibitem[Macci\`o et al. (2010)]{maccio09} Macci\`o A. V., Kang, X., Fontanot, F., et al. 2010, \mnras, 402, 1995

\bibitem[Madau et al. (2008)]{madau08} Madau, P., Diemand, J., \& Kuhlen, M. 2008, \apj, 679, 1260

\bibitem[Moore et al. (1999)]{moore99} Moore, B., Ghigna, S., Governato, F., et al. 1999, 524, L19

\bibitem[Navarro et al. (1997)]{nfw} Navarro, J. F., Frenk, C. S., \& White S. D. M. 1997, \apj, 490, 493

\bibitem[Navarro et al. (2004)]{navarro04} Navarro, J. F., Hayashi, E., Power, C., et al. 2004, \mnras, 349, 1039

\bibitem[Navarro et al. (2010)]{navarro10} Navarro, J. F., Ludlow, A., Springel, V., et al. 2010 \mnras, 402, 21

\bibitem[Neto et al. (2007)]{neto07} Neto, A. F., Gao, L., Bett, P., et al. 2007, MNRAS, 381, 1450

\bibitem[Odenkirchen et al. (2001)]{pal5} Odenkirchen, M., Grebel, E. K., Rockosi, C. M., et al. 2001, \apjl, 548, L165

\bibitem[Odenkirchen et al. (2009)]{odenkirchen09} Odenkirchen, M., Grebel, E. K., Kayser, A., et al. 2009, \aj, 137, 3378

\bibitem[Perlmutter et al. (1999)]{perlmutter99} Perlmutter, S., Aldering, G., Goldhaber, G., et al. 1999, \apj, 517, 565

\bibitem[Planck Collaboration et al. (2013)]{planck13} Planck Collaboration, et al. 2013, arXiv:1303.5062

\bibitem[Riess et al. (1998)]{riess98} Riess, A. G., Filippenko, A. V., Challis, P., et al. 1998, \aj, 116, 1009

\bibitem[Schneider et al. (2013)]{schneider13} Schneider, A., Smith, R. E., \& Reed, D. 2013, \mnras, 433, 1573

\bibitem[Siegal-Gaskins \& Valluri (2008)]{SV08} Siegal-Gaskins, J. M., \& Valluri, M. 2008, \apj, 681, 40

\bibitem[Spergel \& Steinhardt (2000)]{spergel00} Spergel D. N., \& Steinhardt, P. J. 2000, PhRvL, 84, 3760

\bibitem[Springel (2005)]{gadget2} Springel, V. 2005, MNRAS, 364, 1105

\bibitem[Springel et al.(2008)]{aquarius} Springel, V., Wang, J., Vogelsberger, M., et al. 2008, \mnras, 391, 1685

\bibitem[Stadel et al. (2009)]{stadel09} Stadel, J., Potter, D., Moore, B., et al. 2009, \mnras, 398, L21

\bibitem[Strigari et al. (2007)]{strigari07} Strigari, L. E., Bullock, J. S., Kaplinghat, M. 2007, \apj, 669, 676

\bibitem[Strigari et al. (2008)]{strigari08} Strigari, L. E., Bullock, J. S., Kaplinghat, M., et al. 2008, Nature, 454, 1096

\bibitem[Willett et al. (2009)]{willett09} Willett, B. A., Newberg, H. Jo, Zhang, H., et al. 2009, \apj, 697, 207

\bibitem[Yoon et al.(2011)]{yoon11} Yoon, J. H., Johnston, K. V., \& Hogg, D. W. 2011, \apj, 731, 58

\bibitem[Zemp et al. (2009)]{zemp09} Zemp, M., Diemand, J., Kuhlen, M., et al. 2009, \mnras, 394, 641

\end{thebibliography}
\end{document}